\documentclass[superscriptaddress,prx, longbibliography]{revtex4-2}
\usepackage[utf8]{inputenc}
\usepackage{physics}
\usepackage{amsmath,amsfonts}
\usepackage{indentfirst}
\usepackage{array}
\usepackage{amsthm}
\usepackage{amssymb}
\usepackage{geometry}
\usepackage{float}
\usepackage{makecell,xcolor,soul}

\usepackage[inkscapearea=page]{svg}
\usepackage{adjustbox}
\usepackage{hyperref}

\newcommand{\mh}[1]{\textcolor{black}{#1}}
\newcommand{\yc}[1]{\textcolor{black}{#1}}

\usepackage{chemformula} %% $\ch{H2O}$ etc
\usepackage{qcircuit}
\usepackage{booktabs}
\usepackage{subcaption}
\usepackage{cleveref}

\begin{document}

%\title{Quantum Advantage in Small Molecule Drug Discovery}
\title{Exploring the Advantages of Quantum Generative Adversarial Networks in Generative Chemistry}

\author{Po-Yu Kao}
\affiliation{Insilico Medicine Taiwan Ltd., Taipei, Taiwan}

\author{Ya-Chu Yang}
\affiliation{Insilico Medicine Taiwan Ltd., Taipei, Taiwan}

\author{Wei-Yin Chiang}
\affiliation{Hon Hai (Foxconn) Research Institute, Taipei, Taiwan}

\author{Jen-Yueh Hsiao}
\affiliation{Hon Hai (Foxconn) Research Institute, Taipei, Taiwan}

\author{Yudong Cao}
\affiliation{Zapata Computing, Inc., Boston, Massachusetts, USA}

\author{Alex Aliper}
\affiliation{Insilico Medicine AI Limited, Masdar City, Abu Dhabi, UAE}

\author{Feng Ren}
\affiliation{Insilico Medicine Shanghai Ltd., Shanghai, China}

\author{Alán Aspuru-Guzik}
\affiliation{Department of Chemistry, University of Toronto, Toronto, Canada}
\affiliation{Department of Computer Science, University of Toronto, Toronto, Canada}
\affiliation{Vector Institute for Artificial Intelligence, Toronto, Canada}
\affiliation{Lebovic Fellow, Canadian Institute for Advanced Research, Toronto, Canada}

\author{Alex Zhavoronkov}
\thanks{Corresponding e-mail: alex@insilico.com}
\affiliation{Insilico Medicine Hong Kong Ltd., Hong Kong SAR, China}

\author{Min-Hsiu Hsieh}
\thanks{Corresponding e-mail: min-hsiu.hsieh@foxconn.com}
\affiliation{Hon Hai (Foxconn) Research Institute, Taipei, Taiwan}

\author{Yen-Chu Lin}
\thanks{Corresponding e-mail: jimmy.lin@insilico.com}
\affiliation{Insilico Medicine Taiwan Ltd., Taipei, Taiwan}
\affiliation{Department of Pharmacy, National Yang Ming Chiao Tung University, Taipei, Taiwan}

\date{\today}

\maketitle

\section*{Abstract}

\textit{De novo} drug design with desired biological activities is crucial for developing novel therapeutics for patients. 
The drug development process is time and resource-consuming, and it has a low probability of success. 
Recent advances in machine learning and deep learning technology have reduced the time and cost of the discovery process and therefore, improved pharmaceutical research and development.  
In this paper, we explore the combination of two rapidly-developing fields with lead candidate discovery in the drug development process.
First, Artificial intelligence has already been demonstrated to successfully accelerate conventional drug design approaches. 
Second, quantum computing has demonstrated promising potential in different applications, such as quantum chemistry, combinatorial optimizations, and machine learning. 
This manuscript explores hybrid quantum-classical generative adversarial networks (GAN) for small molecule discovery. 
We substituted each element of GAN with a variational quantum circuit (VQC) and demonstrated the quantum advantages in the small drug discovery. 
Utilizing a VQC in the noise generator of a GAN to generate small molecules achieves better physicochemical properties and performance in the goal-directed benchmark than the classical counterpart. 
Moreover, we demonstrate the potential of a VQC with only tens of learnable parameters in the generator of GAN to generate small molecules.
We also demonstrate the quantum advantage of a VQC in the discriminator of GAN. 
In this hybrid model, the number of learnable parameters is significantly less than the classical ones, and it can still generate valid molecules. 
The hybrid model with only tens of training parameters in the quantum discriminator outperforms the MLP-based one in terms of both generated molecule properties and the achieved KL divergence.

\section{Introduction}

% overview of drug discovery
The drug development process includes discovery and development, pre-clinical research, clinical research, Food and Drug Administration (FDA) review, and FDA post-market safe monitoring.
The entire process is time and resource-consuming and has a low probability of success with approximately 4\% of pre-clinical drugs eventually granted license \cite{hingorani2019improving}.
The average time for a new medicine to complete the journey from initial discovery to the marketplace takes at least ten years \cite{abreu2020ivermectin}.
The estimated median capitalized research and design (R\&D) cost per new drug (accounting for the cost of failures) was \$985 million between 2009 and 2018 \cite{wouters2020estimated}.
Recent advances in machine learning and deep learning technology have improved and reduced the cost of pharmaceutical R\&D \cite{aliper2016deep,kadurin2017cornucopia,zhavoronkov2018artificial,carracedo2021review,kao2021toward,kolluri2022machine}. 
For example, some of us \cite{zhavoronkov2019deep} discovered potent inhibitors of discoidin domain receptor 1 (DDR1), a kinase target implicated in fibrosis and other diseases, in 21 days. 
Chan et al. \cite{chan2019advancing} estimated their machine learning algorithms would shrink the drug candidate identification phase from a few months to one year. 

\textit{De novo} drug design refers to a novel chemical compound design with desired pharmacological and physicochemical properties \cite{schneider2005computer}.
The discovery of novel chemical compounds with desired biological activities is a critical step to keep the drug discovery pipeline moving forward \cite{fischer2019approaching}. 
It is also crucial for developing novel therapeutics for patients \cite{mouchlis2021advances}.
Conventional approaches include ligand-based drug design (LBDD), fragment-based drug design (FBDD), and structure-based drug design (SBDD).
LBDD is based on known active binders of a biological target, and FBDD identifies small molecular fragments with weak affinity for a biomolecular target of interest and assembles them into fully drug-like compounds \cite{speck2018recent}.
Aside from LBDD and FBDD, SBDD is based on the properties of the active site of a biological target.

Artificial intelligence (AI) has made a breakthrough in the recent \textit{de novo} molecule design \cite{mamoshina2018converging,zhavoronkov2020potential,paul2021artificial}.
We identify four well-known generative machine learning algorithms in the field: evolutionary algorithms (EA), recurrent neural network (RNN) such as gated recurrent unit (GRU) and long short-term memory (LSTM), autoencoders such as adversarial autoencoder (AAE) and variational autoencoder (VAE), and generative adversarial network (GAN) \cite{martinelli2022generative}.
GAN \cite{goodfellow2014generative} has become a popular network architecture for generating highly realistic data \cite{karras2017progressive}, and it has shown remarkable results for generating data that mimics a data distribution in different tasks \cite{isola2017image,zhu2017unpaired,karras2019style,jangid20223d}.
GAN consists of a generator and a discriminator defined by an artificial neural network (ANN).
The parameters of a GAN can be learned by backpropagation.
The generator takes random noises as input and tries to imitate the data distribution, and the discriminator tries to distinguish between the fake and real samples. 
A GAN is trained until the discriminator cannot distinguish the generated data from the real data.

% introduce to the previous GANs in molecule generations
\mh{GANs are one of the most successful generative models in the drug discovery field, and }several different GAN architectures have been proposed in the past decades in \textit{de novo} drug discovery \cite{kadurin2017drugan,vanhaelen2020advent,xu2021deepgan}. 
\mh{Zhavoronkov et al. \cite{zhavoronkov2019deep} proposed a deep generative model called generative tensorial reinforcement learning (GENTRL) for \textit{de novo} small molecule generation. 
The GENTRL generates novel drugs with better synthetic feasibility and biological activity.}
Guimaraes et al. \cite{guimaraes2017objective} proposed an objective-reinforced generative adversarial network (ORGAN) that combines the GAN and reinforcement learning (RL) algorithm.
ORGAN is built on SeqGAN \cite{yu2017seqgan} and is the first GAN architecture in the \textit{de novo} molecule generation.
It is a sequential generative model operating on simplified molecular-input line-entry system (SMILES) string representations of molecules. 
The generated samples of ORGAN maintain information originally learned from data, retain sample diversity, and show improvement in the desired drug properties.
Prykhodko et al. \cite{prykhodko2019novo} presented a novel neural network architecture called LatentGAN for \textit{de novo} molecular design. 
It combines an autoencoder and a generative adversarial neural network.
The generator and discriminator of LatentGAN take $n$-dimensional vectors as inputs.
These inputs derived from the code layer of an autoencoder are trained as a SMILES heteroencoder \cite{kotsias2020direct}.
This method allows LatentGAN to focus on optimizing the sampling without worrying about SMILES syntax issues.
Cao and Kipf introduced MolGAN \cite{de2018molgan} for small molecule \textit{de novo} design which operates directly on graph-structured data.
It is the first GAN to address the generation of graph-structured data in the context of molecular generation. 
MolGAN is demonstrated to generate close to 100\% valid compounds in experiments on the quantum machines 9 (QM9) chemical database. 
The generated molecules of MolGAN have better chemical properties particularly in synthesizability and solubility than the generated compounds of ORGAN.
Neither ORGAN nor MolGAN is directly compared with LatentGAN in the paper. 
\mh{These generative models have the potential to be improved with quantum machine learning algorithms.}

AI-driven strategies have the potential to explore adjacent chemical space compared to classical discovery efforts in \textit{de novo} drug design  \cite{jayatunga2022ai}.
However, the traditional generative machine learning algorithms have difficulty exploring the new chemical space different from the training dataset.
\yc{The Hilbert space of the molecule scales exponentially with molecule size.}
%The state space dimension that represents molecule structure scales quadratically with molecule size. 
It increases the difficulty for the classical generative model on sampling as the landscape becomes extremely large and impossible to screen all possible configurations.
Thus, an efficient sampling method is needed. 

The applications of quantum computing can be found in different fields, such as solving routing problems \cite{neukart2017traffic, harwood2021formulating}, stock price forecasting  \cite{orus2019quantum,liu2022quantum}, multi-task classification \cite{du2022demystify},  two-player zero-sum game \cite{yin2022efficient}, \yc{high-resolution handwritten digits generation \cite{rudolph2022generation},} the discovery of molecular properties  \cite{aspuru2005simulated,lanyon2010towards,peruzzo2014variational,o2016scalable,kandala2017hardware,cao2019quantum,google2020hartree, gao2021computational}, tautomeric state prediction \cite{shee2022quantum}, and drug design \cite{cao2018potential,li2021quantum}.  We can obtain more efficient and accurate results by utilizing the technique of quantum annealer, quantum machine learning, and quantum algorithm. These advantages that utilize the fundamental properties of quantum mechanics to achieve better performances compared to the classical methods are called quantum advantages. These performances can be on the computational resources requirement in simulation, computational efficiency in algorithm development, and measurement precision in metrology. More specifically, quantum advantages can be identified in three aspects: scalability, complexity, and accuracy. The idea of utilizing the intrinsic quantum properties of quantum processors to simulate quantum systems was proposed by Feynman in 1982 \cite{feynman1982simulating}, allowing us to understand the characteristics of large quantum systems with a feasible computational resource and achieving the quantum advantage of scalability. For complexity, researchers found that the complexity of many algorithms  \cite{grover1996fast, coppersmith2002approximate,shor1994algorithms,kitaev1995quantum} can be significantly reduced by the quantum properties. Better measurement accuracy can be achieved by sophisticated designed protocols that enhance the accuracy from the standard quantum limit to the Heisenberg limit \cite{giovannetti2004quantum,giovannetti2006quantum,lloyd2008enhanced,komar2014quantum}

Although quantum computing seems to lead a significant leap in solving resource-demanding problems, a few challenges remain to be overcome, such as the imperfect reliability of the qubits prohibits the physical implementation of large-scale quantum computers   \cite{chao2018fault}. Nevertheless, noisy intermediate-scale quantum (NISQ) computers can still perform classical challenge tasks accompanied by quantum advantages  \cite{bharti2022noisy}. For example, the advantages are shown in sampling the output of a pseudo-random quantum circuit \cite{arute2019quantum}, the sampling time complexity of a Torontonian of a matrix \cite{zhong2020quantum, quesada2018gaussian}, expressively in unsupervised learning \cite{gao2021enhancing}, and parameter space complexity in reinforcement learning \cite{chen2020variational}. 

Although the idea of data conversion in the quantum circuit through gate operators is similar to the classical neural network, efficient sampling that can be achieved by quantum circuits as the fundamental principle of quantum mechanics \yc{enables computational processes that are beyond what is possible on classical computers.}
%allows the coexistence of all possible states exponentially large in the quantum circuit. 
%This fundamental property of superposition in quantum mechanics gives exponentially larger state space, allowing efficient sampling that is difficult for the classical model. 
Several studies have demonstrated that variational quantum circuit (VQC) \cite{bharti2022noisy} performs the advantages in expression power \cite{du2020expressive}, learnability \cite{du2021learnability}, and robustness \cite{du2021quantum}. 
It indicates that VQC \yc{can greatly boost the} solution to the problem, e.g., drug discovery, which is hard to tackle by the classical neural network.
In addition, Blunt et al. \cite{blunt2022perspective} discuss that using the current best theoretical algorithms on a quantum computer would reduce the pharmaceutical R\&D period from many years to only a week.

%Successful cases of Qml in drug discovery and molecule structure prediction
The innate advantages of VQC are suited for computational resource-demanding tasks, such as identifying possible chemical reaction pathways and predicting molecule structures. 
The ground state energy of different molecule configurations reveals the most plausible reaction mechanism. 
It is demonstrated that the ground state energy of small molecules, such as $\text{H}_2$, $\text{LiH}$, $\text{BeH}_2$ \cite{kandala2017hardware}, $\text{H}_{12}$, and diazene \cite{google2020hartree}, can be obtained by VQC, \yc{where} the parameterized angles in the circuit are updated \yc{iteratively on the classical computer until} the minimum energies are achieved under the given molecule configuration. 
Furthermore, the obtained ground state energies are shown to be consistent with the results from the classical Full-CI calculation \cite{gao2021computational}. 
In addition, Gircha et al. \cite{gircha2021training} trained a discrete variational autoencoder (DVAE) for generative chemistry and drug design.
Their model is small enough to fit an annealer and produce drug-like molecules.  

Quantum generative adversarial network (QuGAN) \cite{lloyd2018quantum} provided the first theoretical framework of quantum adversarial learning. QuGAN's exponential advantages over classical GANs directly result from the ability of quantum information processors to represent $N$-dimensional features using $log N$ qubits with a time complexity of $O(poly(\operatorname{log}N))$. Recent studies also showed that generative models implemented by quantum circuits with fewer architectural complexities could easily bypass their classical counterparts \cite{huang2021experimental,li2021quantum}.
Dallaire-Demers et al. provided the first feasible implementation of QuGAN using quantum circuits in a simulator \cite{dallaire2018quantum}. 
\yc{Analogous proposals of QuGAN for continuous functions have been proposed during the same period \cite{romero2021variational}.}
Later on, QuGANs demonstrated its first successful training on the MNIST dataset on a physical quantum device \cite{huang2021experimental}. Li et al. \cite{li2021quantum} pushed the research to the real world further by showing the QuGANs can learn or generate the distribution of the QM9 dataset, which provides the quantum chemical properties for small organic molecules in drug design.
%However, the source code \footnote{\url{https://github.com/jundeli/quantum-gan}} they provided struggles with generating training-set-like molecules. 
However, the source code (\url{https://github.com/jundeli/quantum-gan}) they provided struggles with generating training-set-like molecules. 
In addition, it lacks a detailed comparison between the generated samples from QuGAN and those from classical GAN.

% our contribution
In this work, we perform the training tasks on the QM9 dataset using the classical and quantum GAN.
We not only demonstrate that the quantum GAN outperforms the classical GAN in the drug properties of generated compounds and the goal-directed benchmark but ensure that the trained quantum GANs can generate training-set-like molecules by using the variational quantum circuit as the noise generator. 
In addition, we show the potential of the variational quantum circuit in the generator of GAN to generate small molecules. 
In the end, we demonstrate that the quantum discriminator of GAN outperforms the classical counterpart in terms of generated molecule properties and KL-divergence score. 
The source codes will be publicly available once this manuscript is accepted.
\mh{In addition, the hybrid models are planned to be integrated into the Insilico Medicine Chemistry42\textsuperscript{TM} \cite{ivanenkov2021chemistry42} shortly. }

\section{Methodology}

Generative adversarial net (GAN) \cite{goodfellow2014generative} was first proposed in 2014.
It consists of two elements: a generator and a discriminator. 
The objective of the generator is to generate fake data which mimics the real training data, and the goal of the discriminator is to distinguish real data from fake data.
To achieve these goals, the generator and discriminator are trained at the same time using the training data. 
During the training process, both elements compete with each other and iteratively improve with one another.
By the end, the generator can generate novel data highly similar to the data from the training dataset.
More details of GAN can be found in the original GAN \cite{goodfellow2014generative} paper.
Quantum generative adversarial networks (QuGANs) \cite{dallaire2018quantum,lloyd2018quantum} were first introduced around late July 2018. 
Dallaire-Demers and Killoran \cite{dallaire2018quantum} extended adversarial training to the quantum domain, built generative adversarial networks using quantum circuits, and demonstrated that QuGANs can be trained successfully.
\yc{Romero and Aspuru-Guzik proposed analogous construction of QuGAN for learning continuous functions \cite{romero2021variational}.}
Lloyd and Weedbrook \cite{lloyd2018quantum} demonstrated that when the data consists of samples of measurements made on high-dimensional spaces, QuGANs may exhibit an exponential advantage over classical GANs.

In the classical GAN \cite{goodfellow2014generative} model, the inputs to the generator are randomly sampled from a distribution, e.g., uniform distribution or Gaussian distribution, and the generator and discriminator consist of neural networks. 
To demonstrate the quantum advantage of different components of GAN in small drug discovery, we alter each component into a variational quantum circuit (VQC) step by step. 
In the first experiment, we replace the sampling part of GAN with a VQC to generate the noise for the generator. 
In the second experiment, we substitute the classical generator of GAN with a VQC.
This quantum generator takes the noises from the Gaussian distribution as input and outputs the molecular graph to the classical discriminator.
In the last experiment, we replace the classical discriminator with a VQC. 
This quantum discriminator takes the molecular graph as input and predicts if the molecules are real or fake using the measurement from only one qubit. 

In this work, MolGAN \cite{de2018molgan} is used as the base model.
MolGAN with VQC (QuMolGAN) as the noise generator, MolGAN with a quantum generator \cite{huang2021experimental} (MolGAN-QC), and MolGAN with a quantum discriminator (MolGAN-CQ) are used to demonstrate the quantum advantage in the small molecule drug discovery.
All the combinations of the classical/quantum noise/generator/discriminator and their corresponding model name are shown in \Cref{tbl:model_name}.
The whole pipeline is implemented by using Pennylane \cite{bergholm2018pennylane} and PyTorch \cite{paszke2017automatic}.  
The details of VQC, VQC of the noise generator, VQC of the generator, and VQC of the discriminator will be introduced as follows. 

\begin{table}[hbt!]
\caption{All the combinations of the classical/quantum noise/generator/discriminator and their corresponding model name.}
\label{tbl:model_name}
\begin{tabular}{llll}
\hline
\textbf{Model Name} & \textbf{Noise} & \textbf{Generator} & \textbf{Discriminator} \\ \hline
MolGAN \cite{de2018molgan}& classical & classical          & classical graph-based \\ 
QuMolGAN            & quantum   & classical          & classical graph-based \\ 
MolGAN-QC           & classical & quantum            & classical graph-based \\
MolGAN-CQ           & classical & classical          & quantum \\ 
MolGAN-CC           & classical & classical & classical MLP-based   \\ \hline
\end{tabular}
\end{table}

\subsection{Variational Quantum Circuits (VQCs)}

A Quantum gate is a basic quantum circuit operating on a small number of qubits.
In this work, two controlled gates (Controlled-X gate and Controlled-Z gate) and four \yc{single-qubit rotations} ($R_x$, $R_y$, $R_z$, and $R$) are used to construct the variational quantum circuit. 
Controlled-X (CNOT) gate is a two-qubit operation, where the first qubit is usually referred to as the control qubit and the second qubit as the target qubit. 
\begin{equation}
CNOT = 
\begin{bmatrix}
1 & 0 & 0 & 0 \\
0 & 1 & 0 & 0 \\
0 & 0 & 0 & 1 \\
0 & 0 & 1 & 0 \\
\end{bmatrix}
\end{equation}
A Controlled-Z (CZ) gate is a two-qubit operation defined as:
\begin{equation}
CZ = 
\begin{bmatrix}
1 & 0 & 0 & 0 \\
0 & 1 & 0 & 0 \\
0 & 0 & 1 & 0 \\
0 & 0 & 0 & -1 \\
\end{bmatrix}
\end{equation}
The $R_x$, $R_y$, and $R_z$ gates are the essential rotation operators in the quantum circuit. 
The $R_x$ gate is a single-qubit rotation through an angle $\theta$ in radians around the x-axis, and it is defined as:
\begin{equation}
R_x(\theta) = 
\begin{pmatrix}
cos(\theta/2) & -isin(\theta/2)\\
-isin(\theta/2) & cos(\theta/2)
\end{pmatrix}
\end{equation}
The $R_y$ gate is a single-qubit rotation through an angle $\theta$ in radians around the y-axis, and it is defined as:
\begin{equation}
R_y(\theta) = 
\begin{pmatrix}
cos(\theta/2) & -sin(\theta/2)\\
sin(\theta/2) & cos(\theta/2)
\end{pmatrix}
\end{equation}
The $R_z$ gate is a single-qubit rotation through an angle $\theta$ in radians around the z-axis, and it is defined as:
\begin{equation}
R_z(\theta) = 
\begin{pmatrix}
e^{-i\frac{\theta}{2}} & 0 \\
0 & e^{i\frac{\theta}{2}}
\end{pmatrix}
\end{equation}
The $R$ gate is a single-qubit rotation through arbitrary angles {$\alpha$, $\beta$, $\gamma$ } in radians, and it can be decomposed into $R_y$ and $R_z$ gates.
It is defined as:
\begin{equation}
R(\alpha, \beta, \gamma) = R_z(\gamma)R_y(\beta)R_z(\alpha) = 
\begin{pmatrix}
e^{-i\frac{(\alpha+\gamma)}{2}}cos(\beta/2) & -e^{-i\frac{(\alpha-\gamma)}{2}}sin(\beta/2) \\
e^{-i\frac{(\alpha-\gamma)}{2}}sin(\beta/2) & e^{i\frac{(\alpha+\gamma)}{2}}cos(\beta/2)
\end{pmatrix}
\end{equation}
A variational quantum circuit (VQC) shown in \Cref{fig:quantum_circuit_noise_generator,fig:vqc_quantum_generator}consists of three ingredients: (1) the preparation of a fixed initial state, (2) a quantum circuit, and (3) the measurement. 
The initialization layer may contain $R_x$, $R_y$, $R_z$, and $R$ gates, and the rotation angles are sampled from a uniform distribution or Gaussian distribution. 
The parameterized layers which could be repeated for $L$ times could have CNOT gates, CZ gates, and parameterized rotational gates whose parameters (rotation angle) can be learned through the back-propagation. 
The measurement takes the expected value of each qubit.

\subsection{VQC of Noise Generator}

\begin{figure}[hbt!]
    \centering
    %\includesvg[width=0.7\textwidth]{figures/molgan_schema.svg}
    \includegraphics[width=0.7\textwidth]{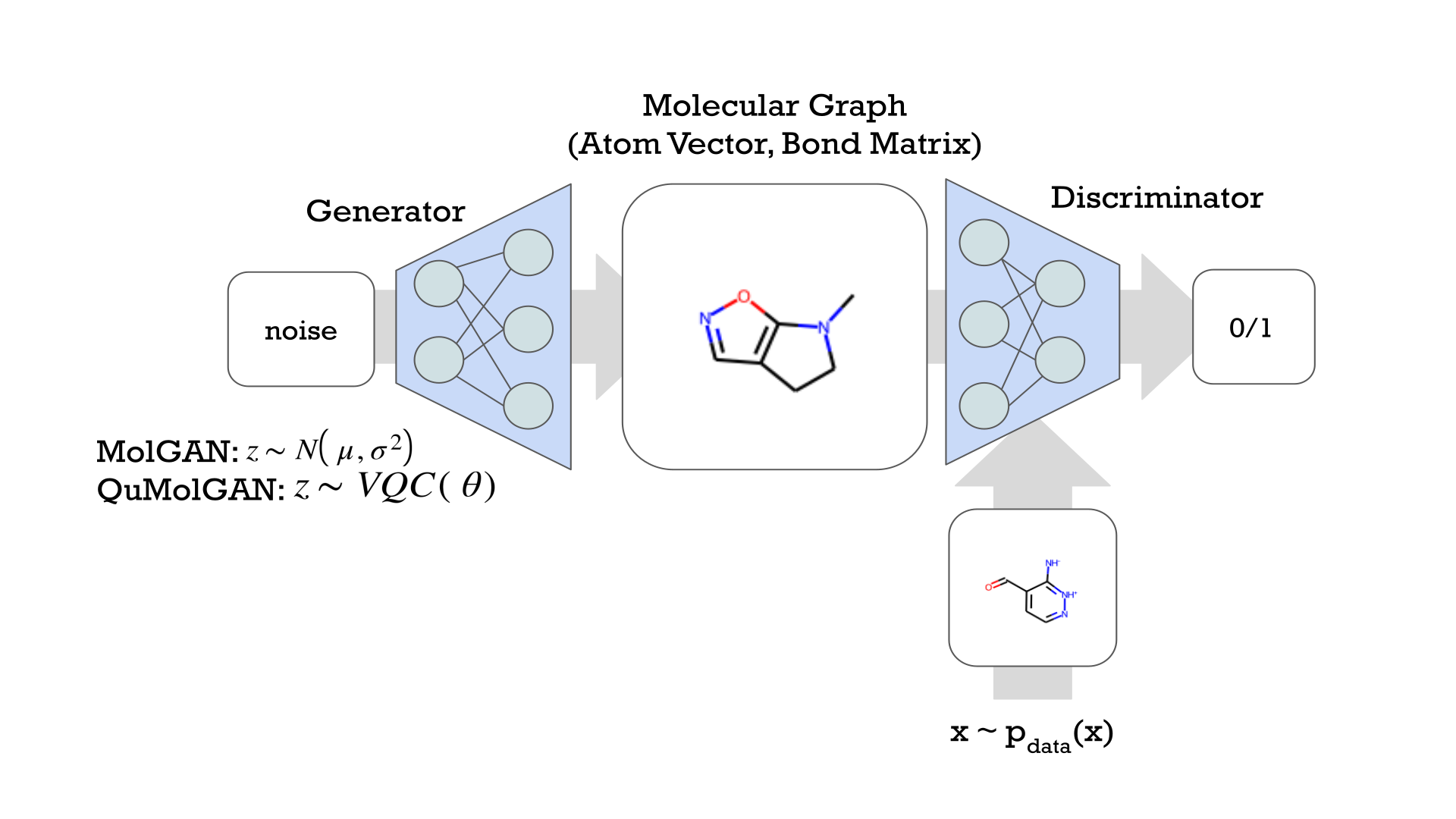}
    \caption{Schema of MolGAN and QuMolGAN. The generator takes the noise as input and generates the molecular graph including the atom vector and bond matrix. The discriminator tries to distinguish between the fake molecular graph from the generator and the real molecular graph from the data distribution.}
    \label{fig:molgan-schema}
\end{figure}

In this work, MolGAN \cite{de2018molgan} is used as the base model for small molecule generation.
We extend its noise generation part to the quantum domain and demonstrate the quantum advantage in the small molecule generation. 
MolGAN \cite{de2018molgan} is an implicit and likelihood-free generative model for the small molecular graph generation.
In contrast to the sequence-based models, MolGAN directly works on the graph representation of molecules.
MolGAN also bypasses the requirement of expensive graph-matching procedures and node-ordering heuristics of likelihood-based methods.
In the classical GANs \cite{goodfellow2014generative}, the inputs to the generator are sampled from a distribution, e.g., uniform distribution or Gaussian distribution. 
Here, we would like to demonstrate the quantum advantage in the small molecule generation by utilizing a variational quantum circuit to generate the inputs of the generator, and we call this hybrid model QuMolGAN.
The schema of MolGAN and QuMolGAN is shown in \Cref{fig:molgan-schema}.

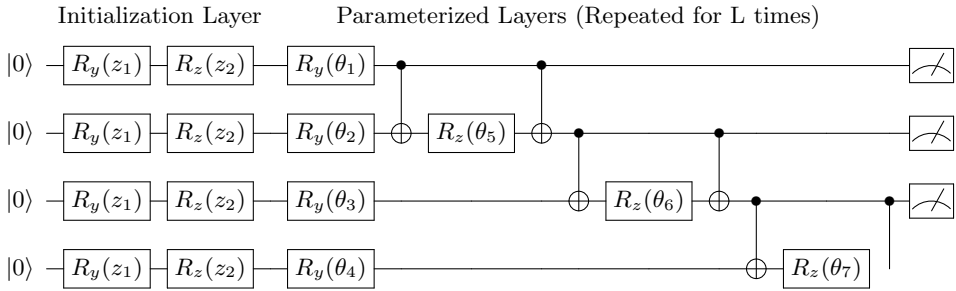
\begin{figure}[hbt!]
\centerline{
\Qcircuit @C=0.7em @R=1.2em {
 &  \mbox{~~~~~~~~~~~~~Initialization Layer} & &&&&&& \mbox{Parameterized Layers (Repeated for L times)} \\
\lstick{\ket{0}} & \gate{R_y(z_1)} & \gate{R_z(z_2)} & \qw & \gate{R_y(\theta_{1})} & \ctrl{1} & \qw                    & \ctrl{1} & \qw      & \qw                    & \qw      & \qw & \qw & \qw & \meter \\
\lstick{\ket{0}} & \gate{R_y(z_1)} & \gate{R_z(z_2)} & \qw & \gate{R_y(\theta_{2})} & \targ    & \gate{R_z(\theta_{5})} & \targ    & \ctrl{1} & \qw                    & \ctrl{1} & \qw & \qw & \qw & \meter \\
\lstick{\ket{0}} & \gate{R_y(z_1)} & \gate{R_z(z_2)} & \qw & \gate{R_y(\theta_{3})} & \qw      & \qw                    & \qw      & \targ    & \gate{R_z(\theta_{6})} & \targ    & \ctrl{1}  & \qw & \ctrl{1} & \meter \\
\lstick{\ket{0}} & \gate{R_y(z_1)} & \gate{R_z(z_2)} & \qw & \gate{R_y(\theta_{4})} & \qw      & \qw                    & \qw      & \qw      & \qw                    & \qw      & \targ     & \gate{R_z(\theta_{7})} &  %\targ &  \meter \gategroup{2}{2}{5}{3}{.7em}{--} \gategroup{2}{5}{5}{14}{.7em}{--} \\
}
}
\caption{The VQC consists of the preparation of the initialization state, a single layer of the 4-qubit ansatz circuit, and the measurement. 
The initialization layer contains $R_y$ and $R_z$ gates, and the rotation angles ($z_1$ and $z_2$) are sampled from a uniform distribution. 
The parameterized layers (could be repeated for $L$ times) have CNOT gate and two types of parameterized rotational gates, $R_y$ and $R_z$ gates whose parameters ($\theta_k$) can be learned through back-propagation. The measurement takes the expected value of each qubit.}
\label{fig:quantum_circuit_noise_generator}
\end{figure}

The generator takes the noise as input and generates the molecular graph including the atom vector and bond matrix. 
The noise is sampled from the Gaussian distribution for the MolGAN, and the noise is generated from the VQC in \Cref{fig:quantum_circuit_noise_generator} for the QuMolGAN.

The discriminator tries to distinguish between the fake molecular graph from the generator and the real molecular graph from the data distribution.
More details of MolGAN can be found in the original MolGAN \cite{de2018molgan} paper.

\subsection{VQC of Quantum Generator}
\begin{figure}[hbt!]
    \centering
    \includegraphics[width=0.75\textwidth]{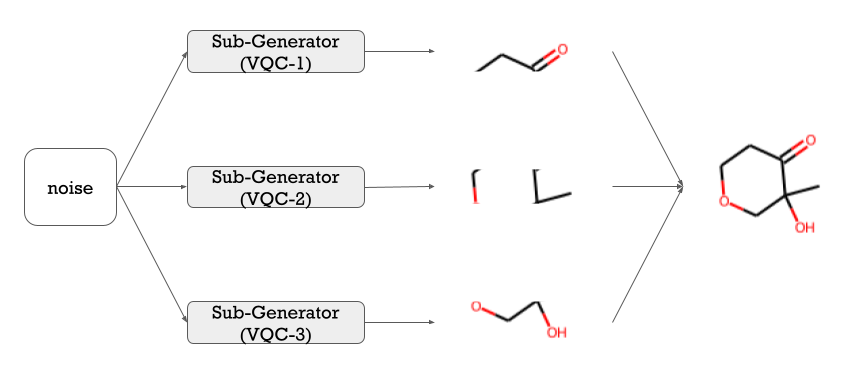}
    \caption{The patch method \cite{huang2021experimental} uses multiple VQCs as sub-generators. Each sub-generator takes noise as input and outputs a partial part of the final molecular graph. The final molecular graph is constructed by concatenating all the partial patches together. }
    \label{fig:patch_method}
\end{figure}

\begin{figure}[hbt!]
\centerline{
\Qcircuit @C=1.4em @R=1.4em {
 &  \mbox{\parbox{4cm}{\noindent Initialization \\ Layer}} && \mbox{\parbox{4cm}{\noindent Parameterized Layers \\ (Repeated for L times)}} \\
\lstick{\ket{0}} & \gate{R_y(z_1)} & \gate{R_y(\theta_{1})} & \ctrl{1} & \qw & \qw & \meter \\
\lstick{\ket{0}} & \gate{R_y(z_2)} & \gate{R_y(\theta_{2})} & \gate{Z} & \ctrl{1} & \qw & \meter \\
\lstick{\ket{0}} & \gate{R_y(z_3)} & \gate{R_y(\theta_{3})} & \qw      & \gate{Z} & \ctrl{1} & \meter \\
\lstick{\ket{0}} & \gate{R_y(z_4)} & \gate{R_y(\theta_{4})} & \qw      & \qw      & \gate{Z} & \meter \gategroup{2}{2}{5}{2}{.7em}{--} \gategroup{2}{3}{5}{6}{.7em}{--} \\
}
}
\caption{The VQC of the quantum generator consists of the preparation of the initialization state, a single layer of the 4-qubit  circuit, and the measurement. 
The initialization layer contains $R_y$ gates, and the rotation angles ($z_i$) are sampled from a uniform distribution. 
The parameterized layers (could be repeated for $L$ times) have CZ gates and one type of parameterized rotational gates, $R_y$ whose parameters ($\theta_k$) can be learned through back-propagation. 
The measurement takes the expected value of each qubit.}
\label{fig:vqc_quantum_generator}
\end{figure}
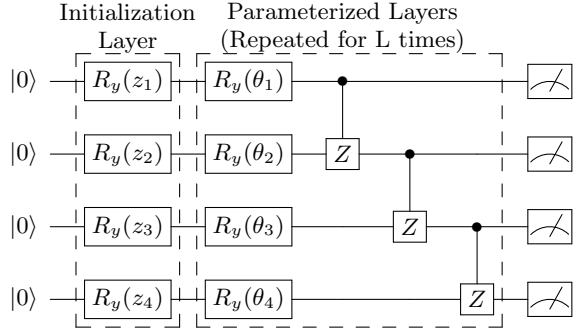

We implement the patch method \cite{huang2021experimental} in the quantum generator of MolGAN (MolGAN-QC). 
This method uses multiple VQCs as sub-generators, and each sub-generator is responsible for constructing a partial part of the final output, e.g., the molecular graph in this study. 
The final molecular graph which consists of the atom vector and bond matrix is constructed by concatenating all the partial patches together as shown in \Cref{fig:patch_method}.
The sub-generator shares the same ansatz architecture as shown in \Cref{fig:vqc_quantum_generator}. 
Each ansatz circuit consists of the preparation of the initialization state, a single layer of a 4-qubit circuit, and the measurement. 
The initialization layer contains $R_y$ gates, and the rotation angles ($z_i$) are sampled from a uniform distribution. 
The parameterized layers (could be repeated for $L$ times) have CZ gates and one type of parameterized rotational gates, $R_y$ whose parameters ($\theta_k$) can be learned through back-propagation.

\subsection{VQC of Quantum Discriminator}

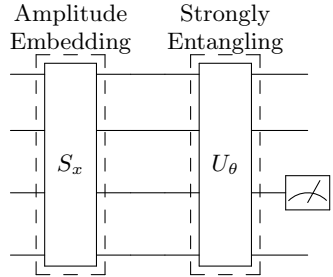
\begin{figure}[hbt!]
\centerline{
\Qcircuit @C=1.4em @R=1.4em {
 &  \mbox{\parbox{4cm}{\noindent Amplitude \\ Embedding}} &&& \mbox{\parbox{4cm}{\noindent Strongly \\ Entangling}} \\
 & \multigate{3}{S_{x}} & \qw & \qw & \multigate{3}{U_{\theta}} & \qw \\
 & \ghost{S_{x}} & \qw &  \qw & \ghost{U_{\theta}} & \qw \\
 & \ghost{S_{x}} & \qw &  \qw & \ghost{U_{\theta}} & \meter \\
 & \ghost{S_{x}} & \qw &  \qw & \ghost{U_{\theta}} & \qw \gategroup{2}{2}{5}{2}{.7em}{--} \gategroup{2}{5}{5}{5}{.7em}{--} \\
}
}
\caption{The VQC of quantum discriminator consists of the amplitude embedding circuit \cite{schuld2018supervised} ($S_x$), the strong entanglement layers \cite{schuld2020circuit} ($U_\theta$), and the measurement. 
The strongly entangling layers \cite{schuld2020circuit} have multiple CNOT gates and parameterized rotational gates ($R$). 
The measurement takes the expectation value of one qubit to determine if the input molecular graph is real or fake.
It is noted that we use 9 qubits to encode the molecular graph in our experiment.}
\label{fig:vqc_quantum_discriminator}
\end{figure}

\begin{figure}[hbt!]
\centerline{
\Qcircuit @C=0.5em @R=0.7em {
& \qw & \gate{R(\alpha_1, \beta_1, \gamma_1)} & \ctrl{1} & \qw       & \qw      & \targ    & \qw & \qw & \gate{R(\alpha_5, \beta_5, \gamma_5)} & \ctrl{2} & \qw      & \targ     & \qw       & \qw & \qw & \qw\\
& \qw & \gate{R(\alpha_2, \beta_2, \gamma_2)} & \targ    &  \ctrl{1} & \qw      & \qw      & \qw & \qw & \gate{R(\alpha_6, \beta_6, \gamma_6)} & \qw      & \ctrl{2} & \qw       & \targ     & \qw & \qw & \qw\\
& \qw & \gate{R(\alpha_3, \beta_3, \gamma_3)} & \qw      &  \targ    & \ctrl{1} & \qw      & \qw & \qw & \gate{R(\alpha_7, \beta_7, \gamma_7)} & \targ    & \qw      & \ctrl{-2} & \qw       & \qw & \qw & \qw\\
& \qw & \gate{R(\alpha_4, \beta_4, \gamma_4)} & \qw      &  \qw      & \targ    & \ctrl{-3}& \qw & \qw & \gate{R(\alpha_8, \beta_8, \gamma_8)} & \qw      & \targ    & \qw       & \ctrl{-2} & \qw & \qw & \qw \gategroup{1}{3}{4}{7}{.7em}{--} \gategroup{1}{10}{4}{15}{.7em}{--} \\
}
}
\caption{The VQC of strongly entanglement layers \cite{schuld2020circuit} contain multiple CNOT gates and parameterized rotational gates ($R$). 
In each layer, each qubit starts with parameterized rotational gates $R(\alpha_i, \beta_i, \gamma_i)$ followed by a CNOT gate.
The parameterized angles $\alpha_i, \beta_i, \gamma_i$ can be learned through back-propagation.
In our experiment, our ansatz structure has 9 qubits and three layers. }
\label{fig:strongly_entangling_layers}
\end{figure}
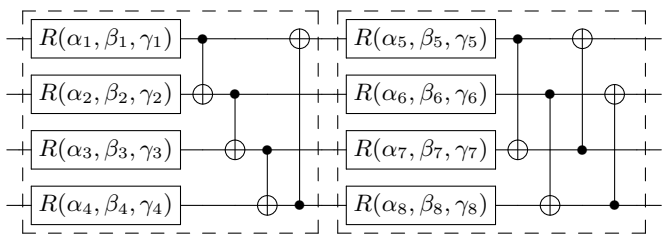

The quantum discriminator takes the molecular graph as input and classifies if this molecular graph is fake (from the generator) or real (from the data distribution).
The VQC of the quantum discriminator consists of the amplitude encoding layer \cite{schuld2018supervised} ($S_x$), the strongly entangling layers ($U_\theta$) inspired by \cite{schuld2020circuit} and the measurement as shown in \Cref{fig:vqc_quantum_discriminator}.
Amplitude encoding is used to encode the atom vector and bond matrix.
The strongly entangling layers \cite{schuld2020circuit} have multiple CNOT gates and parameterized rotational gates $R(\alpha,\beta,\gamma)$ as shown in \Cref{fig:strongly_entangling_layers}. 
In each layer, each qubit starts with parameterized rotational gates $R(\alpha_i, \beta_i, \gamma_i)$ followed by a CNOT gate.
The parameterized angles $\alpha_i, \beta_i, \gamma_i$ can be learned through back-propagation.
The measurement takes the expectation value of one qubit, and this value is used to determine if the input molecular graph is real or fake. 
In our experiment, we use 9 qubits to encode the molecular graph and three-layer of strongly entanglement layers. 

\section{Evaluation Metrics and Dataset}

In this section, we will first introduce the evaluation method and metrics followed by the dataset we used to train the generative models.

\subsection{Evaluation Method and Metrics}

To evaluate the performance of different generative models, we first generate 5,000 noise samples from Gaussian distribution and variational quantum circuits for MolGAN and QuMolGAN, respectively.
The generated noise samples are then fed into the trained generator of models to produce the atom vectors and bond matrices.
In the end, these vectors and matrices are used to construct the molecular graphs. 
Seven evaluation metrics described below will be calculated from the molecular graphs. 

Three quality metrics, i.e., validity, uniqueness, and novelty used in \cite{samanta2018designing,polykovskiy2020molecular}, and three drug properties, i.e., quantitative estimation of drug-likeness (\textit{QED}), solubility, and synthesizability (\textit{SA}), is used to compare different generative models in this work. 
\textit{Validity} is the ratio of the valid molecules to all generated molecules, and \textit{uniqueness} is the ratio of unique molecules to the valid molecules. 
\textit{Novelty} is defined as the ratio of valid molecules which are not in the training dataset to all valid molecules. 
In addition, we also measure the \textit{diversity} of generated molecules which is defined as how likely the generated molecules are to be diverse to the training dataset. 
We also report the valid and unique molecules (\textit{\# molecules}) from 5,000 noise samples. 
\textit{QED} \cite{bickerton2012quantifying} measures how likely a molecule is to be a drug based on the concept of desirability.
\textit{Solubility} reports the n-octanol-water partition coefficient (\textit{logP}) \cite{wildman1999prediction} of the molecule that is the degree of a molecule being hydrophilic.
\textit{SA} \cite{ertl2009estimation} quantifies how easy a molecule is to be synthesized based on the molecular complexity and fragment contributions. 
These property metrics are calculated by using RDKit \cite{greg2022rdkit}.

The Kullback–Leibler (KL) divergences \cite{kullback1951information} are also calculated, and it measures how well a probability distribution approximates another distribution.
The probability distributions of a variety of physicochemical descriptors including BertzCT (molecular complexity index), MolLogP (Wildman-Crippen LogP value \cite{wildman1999prediction}), MolWt (molecular weight), TPSA (molecular polar surface area), NumHAcceptors (number of hydrogen acceptors), NumHDonors (number of hydrogen donors), NumRotatableBonds (number of rotatable bonds), NumAliphaticRings (number of aliphatic rings), and NumAromaticRings (number of aromatic rings) for the generated molecules and the molecules of the training set are compared, and the corresponding KL-divergence scores $D_{KL,i}$ are computed. 
Models able to capture the distributions of molecules in the training set will lead to small KL-divergence scores ($D_{KL}$).
However, the final KL-divergence score ($S$) \cite{brown2019guacamol} used in this paper is computed by 
\begin{equation}
    S = \frac{1}{9}\sum_{i=1}^{9} exp(-D_{KL,i})
\label{eq:kl}
\end{equation}
Therefore, the larger final KL-divergence score (S) means how well the model can capture these nine physicochemical distributions of molecules in the training set.  
\subsection{Dataset}
\begin{figure}[hbt!]
    \centering
    \includegraphics[width=\textwidth]{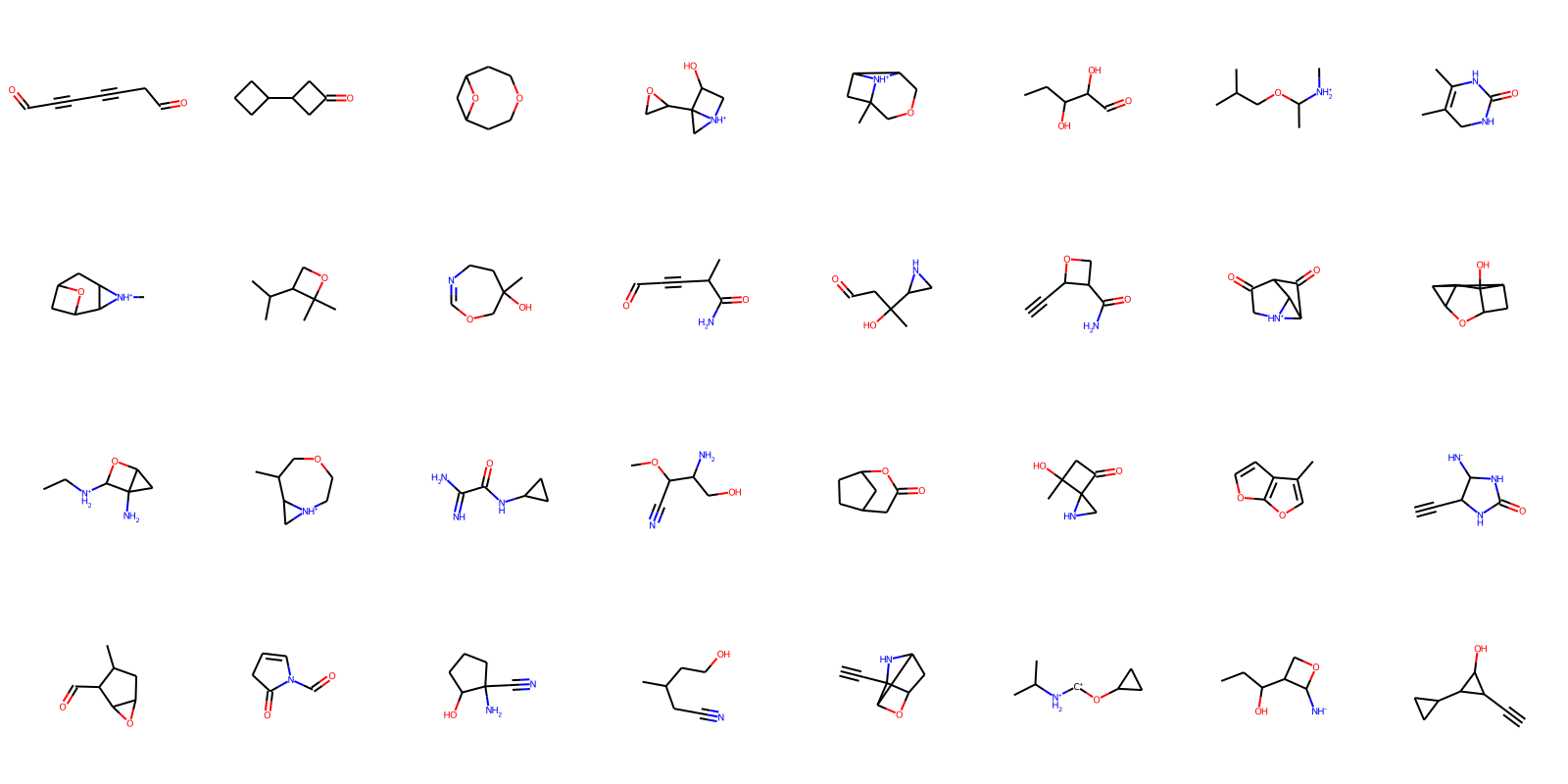}
    \caption{Example molecules of QM9 with canonical SMILES visualized by using RDKit \cite{greg2022rdkit}.}
    \label{fig:qm9_examples}
\end{figure}

All experiments of this work use the QM9 (Quantum Machines 9) \cite{ramakrishnan2014quantum} dataset. 
QM9 dataset is curated from the GDB-17 chemical database \cite{ruddigkeit2012enumeration} which has approximately 166.4 billion molecules. 
QM9 consists of 133,171 molecules containing less than or equal to nine non-hydrogen atoms (carbon, nitrogen, oxygen, and fluorine).  
%The average QED, Solubility, and SA of those 133,171 molecules are 0.461, 0.289, and 0.327, respectively. 
\Cref{fig:qm9_examples} shows some molecules from the QM9 dataset, and \Cref{tbl:qm9_properties} summarizes the drug properties of all molecules of the QM9 dataset.

\begin{table}[hbt!]
\caption{Drug properties of QM9. QED: a quantitative estimate of drug-likeness. Solubility: n-octanol-water partition coefficient (\textit{logP}) \cite{wildman1999prediction}. SA: synthesizability.}
\label{tbl:qm9_properties}
\begin{tabular}{cccc}
\toprule
\multicolumn{4}{c}{QM9}                    \\ \midrule
\# molecules   & QED   & Solubility & SA   \\
133,171        & 0.461 & 0.289  & 0.327 \\ \bottomrule
\end{tabular}
\end{table}

\section{Computational Results and Discussion}

In this section, we would like to computationally explore the potential benefits of our algorithms as compared to classical algorithms in small molecule discovery.
We endeavored to find the best hyper-parameters for the base MolGAN.
We first examined different complexities of generators (in \Cref{appendix:generator_complexity}) and observed that MolGAN-HR (high reduction) has the best performance compared to other generator complexities. 
We then examined different input noise dimensions of generators (in \Cref{appendix:noise_dimension}) and found out that the number of unique and valid molecules saturated at the input dimension equaling 4 (z\_dim=4). 
In addition, we also tested different numbers of parameterized layers in the variational quantum circuit (VQC) (in \Cref{appendix:parameterized_layers}) and observed that MolGAN-HR with 3 parameterized layers of VQC as noise generator has the best performance compared to other numbers of parameterized layers. 
Therefore, in the following experiments, MolGAN-HR is used as the base model, and VQC with 3 parameterized layers is used as the base quantum circuit. 
In the first experiment, we substitute the noise generator of MolGAN with a VQC and discover the quantum advantage in the generated molecules with better drug properties. 
In the second experiment, we replace the generator of MolGAN with a VQC and show the potential of generating small molecules using a VQC.  
In the third experiment, we supplant the discriminator of MolGAN with a VQC and demonstrate the quantum advantage.
All the combinations of the classical/quantum noise/generator/discriminator and their corresponding model name are shown in \Cref{tbl:model_name}.
All experiments are implemented by using Pennylane \cite{bergholm2018pennylane} and PyTorch \cite{paszke2017automatic}.

\subsection{Quantum Noise Generator}

\begin{figure}[htb!] % "[t!]" placement specifier just for this example
\begin{subfigure}{0.33\textwidth}
\includegraphics[width=\linewidth]{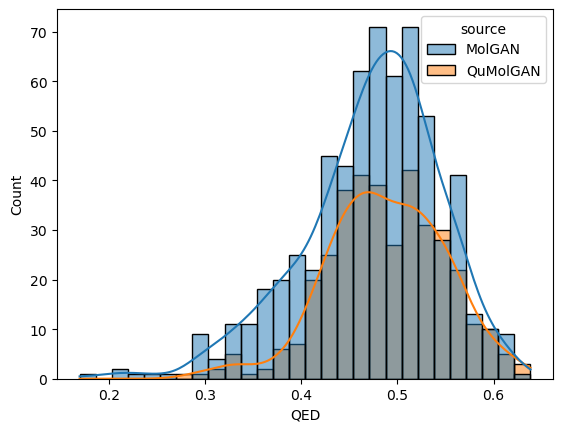}
\caption{QED distribution.} 
\end{subfigure}\hspace*{\fill}
\begin{subfigure}{0.33\textwidth}
\includegraphics[width=\linewidth]{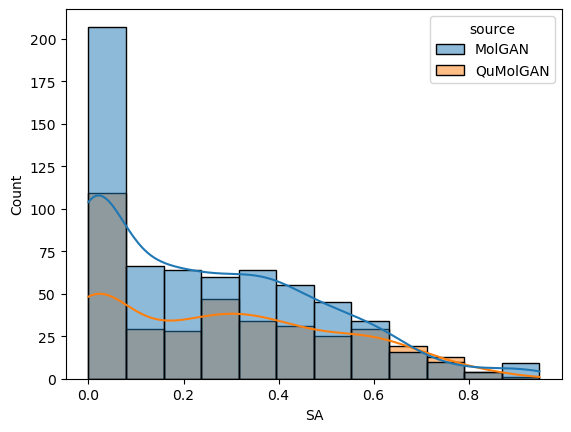}
\caption{SA distribution.} 
\end{subfigure}\hspace*{\fill}
\begin{subfigure}{0.33\textwidth}
\includegraphics[width=\linewidth]{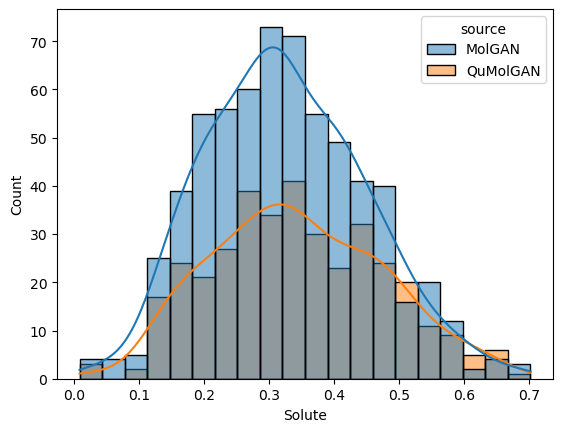}
\caption{Solute distribution.} 
\end{subfigure}
\caption{Distributions of different drug properties from MolGAN-generated and QuMolGAN-generated molecules. From left to right: QED, SA, and Solute distributions. Blue: valid and unique molecules from MolGAN (z\_dim=2). Orange:  valid and unique molecules from QuMolGAN (z\_dim=2).}
\label{fig:drug_properties_dist_molgan_qumolgan}
\end{figure}

\begin{figure}[hbt!]
    \centering
    \includegraphics[width=\textwidth]{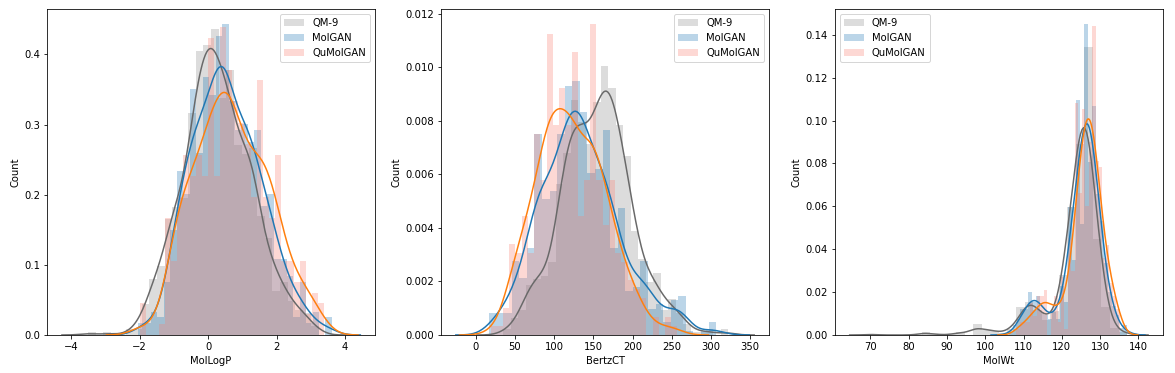}
    \caption{From left to right: the MolLogP, BertzCT, and MolWt distributions of MolGAN-generated and QuMolGAN-generated molecules. Grey: randomly sampled molecules of QM9. Blue: valid and unique molecules from MolGAN (z\_dim=2). Orange:  valid and unique molecules from QuMolGAN (z\_dim=2). }
    \label{fig:kl_distribution_molgan_qumolgan}
\end{figure}

In the first experiment, we would like to compare the performance of QuMolGAN and classical MolGAN. 
We use the same hyper-parameters to train the QuMolGAN and classical MolGAN except for the learning rate.
The learning rate of the quantum noise generator is 0.04, and the learning rate for the generator and discriminator is 0.001.
The models are trained for 150 epochs. 
The WGAN (Wasserstein generative adversarial networks) loss \cite{arjovsky2017wasserstein} is used to train the models. 
QuMolGAN with three parameterized layers is used in this experiment. 
We have examined the input dimension of the generator at 2, 3, 4, and 8, and the results are shown in \Cref{tbl:QuMolGAN_molgan}.
It is noted that the QED, Solute, and SA scores in this table are calculated from the valid and \textbf{\textit{unique}} molecules.
\begin{table}
  \caption{Comparing generated molecules of QuMolGAN and MolGAN in drug properties. Bold numbers highlight the better scores in QuMolGAN compared to the corresponding MolGAN. It is noted that the QED, Solute, and SA scores in this table are calculated from the valid and \textbf{\textit{unique}} molecules.}
  \label{tbl:QuMolGAN_molgan}
  \resizebox{\textwidth}{!}{%
  \begin{tabular}{@{}cccccccccccccccc@{}}
  \toprule
  & \multicolumn{3}{c}{z\_dim=2}        && \multicolumn{3}{c}{z\_dim=3}        && \multicolumn{3}{c}{z\_dim=4}         && \multicolumn{3}{c}{z\_dim=8}        \\ 
  \cmidrule{2-4} \cmidrule{6-8} \cmidrule{10-12} \cmidrule{14-16}
  & QuMolGAN & MolGAN & p-value         && QuMolGAN & MolGAN & p-value         && QuMolGAN & MolGAN & p-value          && QuMolGAN & MolGAN & p-value         \\
  \midrule
  \# molecules\textsuperscript{\emph{a}} & 363      & 657    & -               && 414      & 2163   & -               && 511      & 3085   & -                && 646      & 3287   & -               \\
  QED $\uparrow$          & \textit{\textbf{0.489}} & \textit{\textbf{0.475}} & \textit{\textbf{\textless{}0.01}} && \textit{\textbf{0.489}} & \textit{\textbf{0.465}} & \textit{\textbf{\textless{}0.01}} && \textit{\textbf{0.473}} & \textit{\textbf{0.465}} & \textit{\textbf{\textless{}0.05}}  && 0.474    & 0.470  & 0.149           \\
  Solubility $\uparrow$       & \textit{\textbf{0.343}} & \textit{\textbf{0.324}} & \textit{\textbf{\textless{}0.05}} && \textit{\textbf{0.370}} & \textit{\textbf{0.305}} & \textit{\textbf{\textless{}0.01}} && \textit{\textbf{0.317}} & \textit{\textbf{0.298}} & \textit{\textbf{\textless{}0.01}}  && \textit{\textbf{0.329}} & \textit{\textbf{0.305}} & \textit{\textbf{\textless{}0.01}} \\
  SA $\uparrow$           & \textit{\textbf{0.367}} & \textit{\textbf{0.336}} & \textit{\textbf{\textless{}0.05}} && \textit{\textbf{0.310}} & \textit{\textbf{0.307}} & \textit{\textbf{\textless{}0.05}} && 0.308    & 0.296  & 0.246            && 0.309    & 0.307  & 0.581           \\
  KL Score (S)\textsuperscript{\emph{b}} $\uparrow$ & 0.653 & 0.824 & -  && 0.797 & 0.913 & - && 0.846 & 0.957 & - && 0.868 & 0.957 & - \\ 
  \bottomrule
  \end{tabular}
  }
  \textsuperscript{\emph{a}} Number of valid and unique molecules from 5,000 samples; \\
  \textsuperscript{\emph{b}} From \Cref{eq:kl}; \\
\end{table}
The QuMolGAN is lacking in generating as many molecules as classical MolGAN which results in worse KL Scores.
However, the QuMolGAN can generate molecules with significantly ($p<0.05$) better drug properties compared to the classical MolGAN particularly when the input noise dimension is small (z\_dim=2 and z\_dim=3).
The drug properties distributions of MolGAN-generated and QuMolGAN-generated molecules are shown in \Cref{fig:drug_properties_dist_molgan_qumolgan} for z\_dim=2. 
It shows that QuMolGAN can generate molecules with better drug properties, particularly in QED.
QuMolGAN has less probability to generate molecules whose QED is less than 0.4. 
For the KL-divergence task, the MolLogP, BertzCT, and MolWt distributions of generated molecules are also shown in \Cref{fig:kl_distribution_molgan_qumolgan} for z\_dim=2. 
The classical MolGAN (blue) tends to generate molecules with similar distributions to the training set (grey) which results in a better KL-divergence score.

\begin{figure}[htb!] % "[t!]" placement specifier just for this example
\begin{subfigure}{0.48\textwidth}
%\includesvg[width=\textwidth]{figures/molgan_2z_example_molecules.svg}
\includegraphics[width=\textwidth]{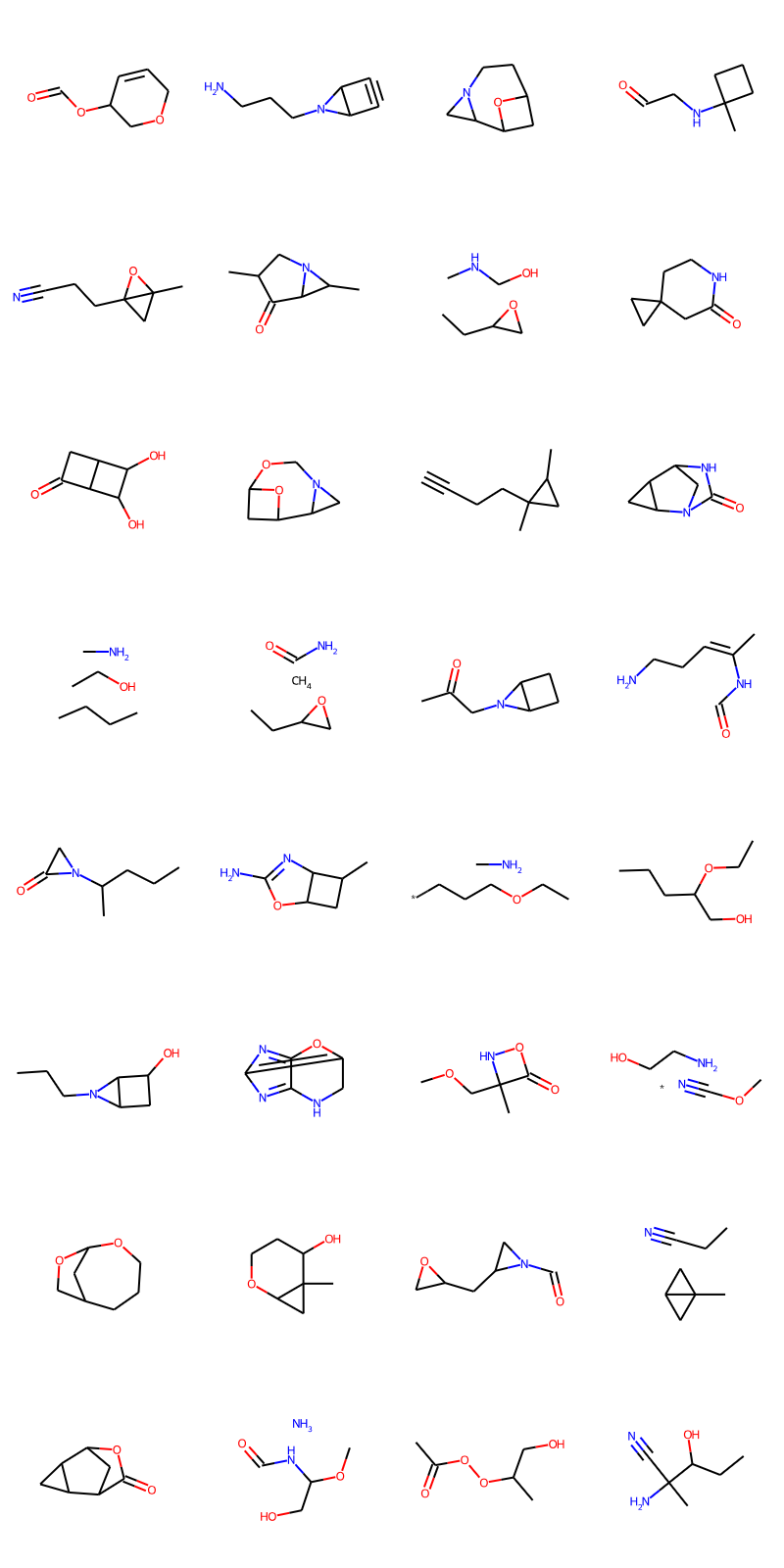}
\caption{Example molecules of MolGAN with z\_dim=2.} 
\end{subfigure}\hspace*{\fill}
\begin{subfigure}{0.48\textwidth}
%\includesvg[width=\textwidth]{figures/quantum_molgan_2z_example_molecules.svg}
\includegraphics[width=\textwidth]{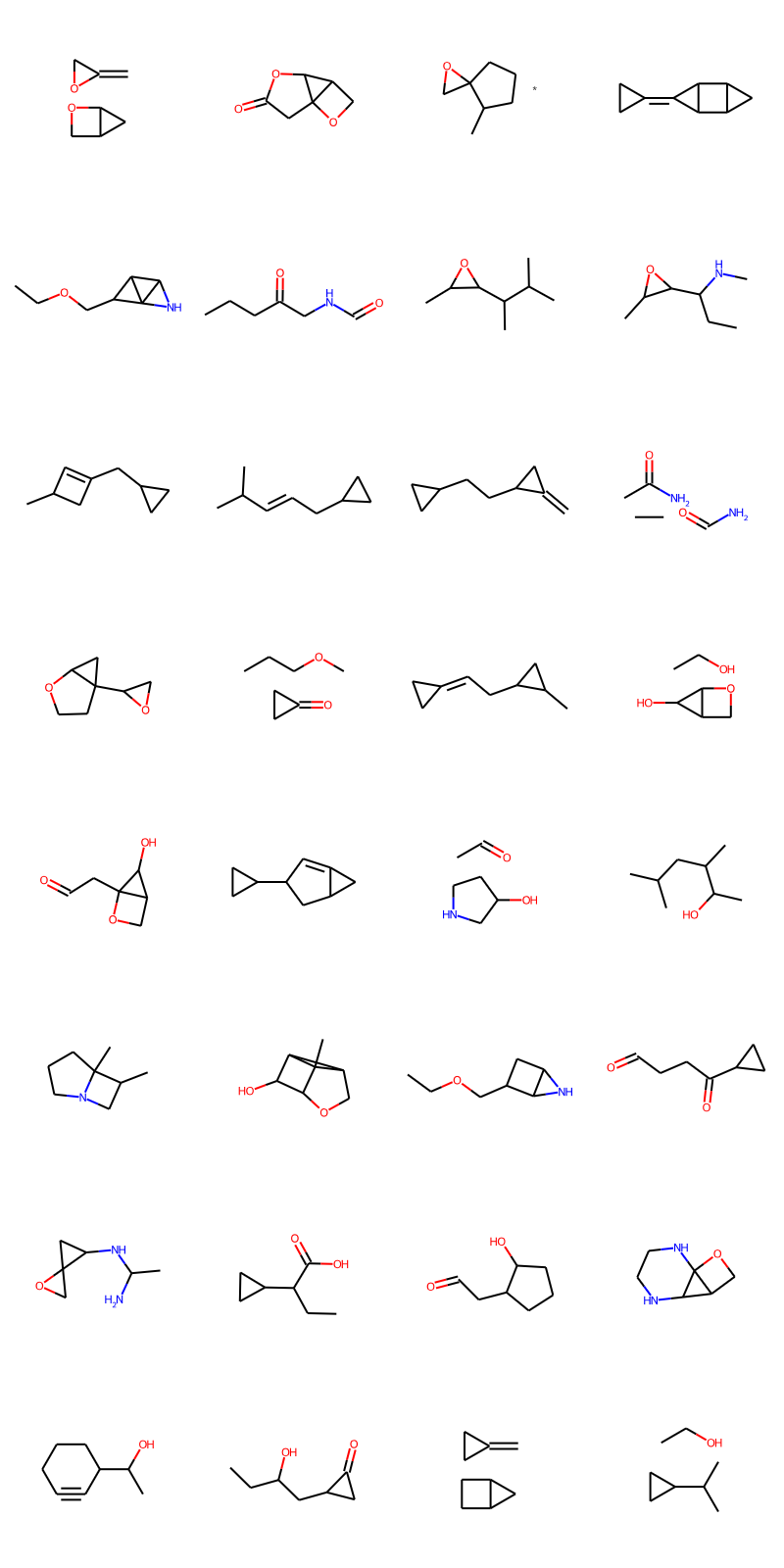}
\caption{Example molecules of QuMolGAN with z\_dim=2.} 
\end{subfigure}
\caption{Example molecules generated by MolGAN (left) and QuMolGAN (right). }
\label{fig:example_molecules}
\end{figure}

In the end, we randomly sample 32 valid and unique molecules from both MolGAN and QuMolGAN for z\_dim=2, and the example molecules are shown in \Cref{fig:example_molecules}.
QuMolGAN can generate training-set-like molecules with better drug properties. 

\begin{figure}[hbt!]
    \centering
    \includegraphics[width=0.7\textwidth]{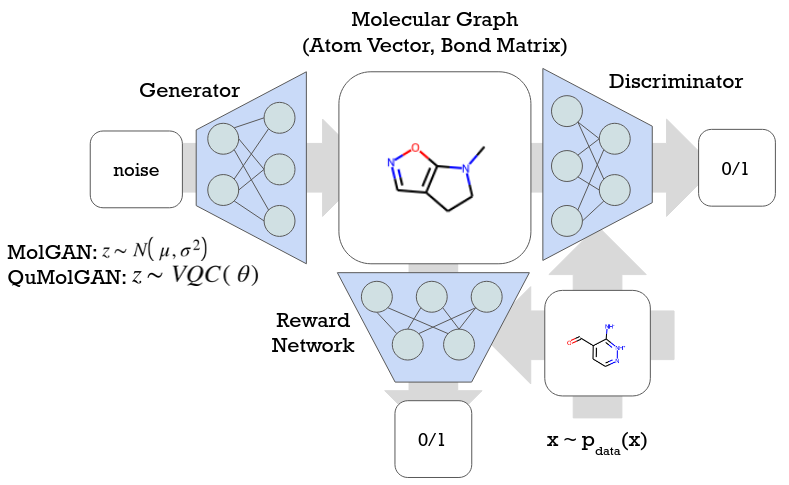}
    \caption{Schema of MolGAN and QuMolGAN with reward network in the goal-directed benchmark. The generator takes the noise as input and generates the molecular graph including the atom vector and bond matrix. The discriminator tries to distinguish between the fake molecular graph from the generator and the real molecular graph from the data distribution. The reward network tries to guide the generator to generate the molecules with desired properties. }
    \label{fig:molgan-reward}
\end{figure}

%\subsubsection{Goal-Directed Benchmark}
As mentioned in \cite{brown2019guacamol}, the goal-directed optimization of molecules tries to improve the demanded scores for the generated molecules.
These scores reflect how well molecules satisfy the required properties. 
The goal is to find molecules that maximize the scoring function.
In this experiment, we also would like to check if the quantum circuit can bring advantages to the MolGAN in the goal-directed benchmark.
Therefore, we add a reward network into the original schema, and the updated schema is shown in \Cref{fig:molgan-reward}. 
At this time, the generator is trained using a linear combination of the WGAN \cite{arjovsky2017wasserstein} loss and the RL \cite{kaelbling1996reinforcement} (reinforcement learning) loss:
\begin{equation}
L(\omega) = \alpha \cdot L_{WGAN}(\omega) + (1-\alpha) \cdot L_{RL}(\omega)
\end{equation}
where $\alpha \in [0, 1] $ is a hyperparameter that controls the trade-off between WGAN loss and RL loss, and $\omega$ are the inputs to networks. 
Here, we set $\alpha = \{0.5, 0.01\}$ to weigh the loss between two components and the goal of improving SA, and QED to train MolGAN and QuMolGAN as suggested by the medical chemists.
In addition, we also add the unique score into the goal to prevent the model from generating the same molecules. 
We have trained MolGAN and QuMolGAN with ($\alpha=\{0.5, 0.01\}$) and without ($\alpha=1.0$) RL loss using the same hyperparameters for 150 epochs, and the results are shown in \Cref{tbl:molgan-rl}.
The input noise dimension to the generator is 4 (z\_dim =4).
It is noted that the QED, Solute, and SA scores in this table are calculated from the valid molecules. 
In this table, the KL Scores of MolGAN are always greater than their quantum counterparts in different weights of RL loss. 
However, QuMolGAN can achieve a higher goal (from 0.47 to 0.57 in QED and from 0.29 to 0.76 in SA) compared to MolGAN (from 0.47 to 0.52 in QED and from 0.30 to 0.60 in SA) while $\alpha=0.01$.
The solute is not the goal so these scores are close for MolGAN and QuMolGAN.  

\begin{table}[hbt!]
\caption{Performance comparison between MolGAN and QuMolGAN with ($\alpha=\{0.5, 0.01\}$) and without ($\alpha=1.0$) reinforcement learning loss in the goal-directed benchmark. It is noted that the QED, Solute, and SA scores in this table are calculated from the valid molecules. Bold numbers indicate better scores among the same type of models with different RL weights $\alpha$, and the underlined numbers indicate the best scores across different types of models.}
\label{tbl:molgan-rl}
\begin{tabular}{ccccccc}
\hline
& \multicolumn{3}{c}{MolGAN} & \multicolumn{3}{c}{QuMolGAN} \\
& $\alpha=1.0$ & $\alpha=0.5$ & $\alpha=0.01$ & $\alpha=1.0$ & $\alpha=0.5$ & $\alpha=0.01$\\ \hline
\#of molecules\textsuperscript{\emph{a}} & \textbf{\underline{2890}} & 2700 & 696 & \textbf{534} & 309 & 116 \\
Validity $\uparrow$ & \textbf{\underline{80.40}} & 78.48 & 68.76 & 70.02 & \textbf{70.32} & 42.94 \\
Uniqueness $\uparrow$ & \textbf{\underline{71.89}} & 68.81 & 20.24 & \textbf{15.25} & 8.78 & 5.40 \\
QED $\uparrow$ & 0.47 & 0.48 & \textbf{0.52} & 0.47 & 0.49 & \textbf{\underline{0.57}} \\
Solute $\uparrow$ & 0.31 & 0.31 & \textbf{\underline{0.45}} & 0.32 & 0.30 & \textbf{0.44}\\
SA $\uparrow$ & 0.30 & 0.31 & \textbf{0.65} & 0.29 & 0.28 & \textbf{\underline{0.76}} \\
KL Score (S)\textsuperscript{\emph{b}}$\uparrow$ & \textbf{\underline{0.95}} & 0.94 & 0.58 & \textbf{0.92} & 0.82 & 0.31\\ \hline    
\end{tabular}\\
\textsuperscript{\emph{a}} Number of valid and unique molecules from 5,000 samples;\\
\textsuperscript{\emph{b}} From \Cref{eq:kl}
\end{table}

\subsection{Quantum Generator}

\begin{figure}[hbt!]
    \centering
    \includegraphics[width=0.7\textwidth]{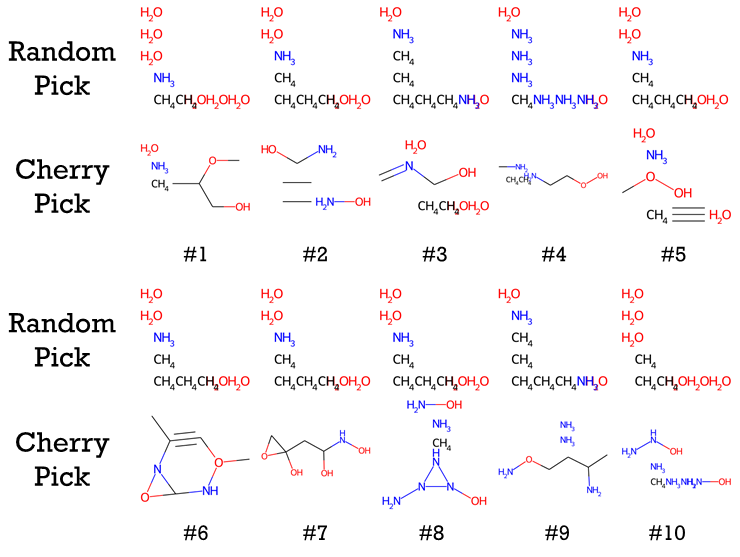}
    \caption{Example molecules from the quantum generator. 
    We input 2048 noises to the quantum generator of MolGAN for every epoch (from \#1 to \#10). 
    The first row shows the random pick of generated molecules, and the second row shows the cherry-pick of generated molecules. }
    \label{fig:quantum_generator_example_molecules}
\end{figure}

In the second experiment, we try to \yc{benchmark} the advantage of the quantum circuit in the generator of GAN. 
We have tried to substitute the generator of MolGAN with a VQC described in \cite{huang2021experimental} for the small molecule generation.
The performance of MolGAN with the quantum generator (MolGAN-QC) is reported in \Cref{tbl:quantum_generator_molgan}.
\begin{table}[hbt!]
  \caption{Performance of MolGAN with the quantum generator. The QED, Solute, and SA scores in this table are calculated from the valid molecules.}
  \label{tbl:quantum_generator_molgan}
  \resizebox{\textwidth}{!}{%
  \begin{tabular}{cccccccccc}
    \toprule 
    \# epoch & \# molecules\textsuperscript{\emph{a}} & validity $\uparrow$ & uniqueness $\uparrow$ & novelty $\uparrow$ & diversity $\uparrow$ & QED $\uparrow$ & Solubility $\uparrow$ & SA $\uparrow$ & KL Score (S)\textsuperscript{\emph{b}} $\uparrow$\\
    \midrule
    1  & 73 & 79.39 & 4.49 & 100 & 1.00 & 0.43 & 0.75 & 0.24 & 0.24 \\
    2  & 54 & 76.37 & 3.45 & 100 & 1.00 & 0.47 & 0.75 & 0.24 & 0.25 \\
    3  & 43 & 78.47 & 2.68 & 100 & 1.00 & 0.48 & 0.75 & 0.11 & 0.28 \\
    4  & 29 & 78.61 & 1.80 & 100 & 1.00 & 0.48 & 0.75 & 0.13 & 0.29 \\
    5  & 30 & 77.93 & 1.88 & 100 & 1.00 & 0.48 & 0.75 & 0.14 & 0.30 \\
    6  & 40 & 78.37 & 2.49 & 100 & 1.00 & 0.48 & 0.75 & 0.09 & 0.21 \\
    7  & 29 & 80.27 & 1.76 & 100 & 1.00 & 0.47 & 0.75 & 0.06 & 0.28 \\
    8  & 39 & 78.91 & 2.41 & 100 & 1.00 & 0.48 & 0.75 & 0.16 & 0.28 \\
    9  & 41 & 74.66 & 2.68 & 100 & 1.00 & 0.48 & 0.75 & 0.22 & 0.25 \\
    10 & 29 & 79.74 & 1.78 & 100 & 1.00 & 0.48 & 0.75 & 0.08 & 0.27 \\
    \bottomrule
  \end{tabular}
  }
  \textsuperscript{\emph{a}} Number of valid and unique molecules from 2,048 samples from Gaussian distribution; \\
  \textsuperscript{\emph{b}} From \Cref{eq:kl}; \\
\end{table}
Although the integration works smoothly, the training processing is time-consuming and resource-consuming. 
The average training time per step takes around 39 seconds in the Amazon EC2 C6a Metal Instance which results in approximately 3.5 days per epoch.
In addition, the model has difficulty generating more valid and unique molecules after being trained for ten epochs.
It fails to generate the training-data-like molecules even after ten epochs of training.
Random-picked and cherry-picked examples of generated molecules are shown in \Cref{fig:quantum_generator_example_molecules}. 
Most generated molecules are similar to the randomly sampled molecules in \Cref{fig:quantum_generator_example_molecules}.
However, we demonstrate that the quantum generator has the potential to generate small molecules.

\subsection{Quantum Discriminator} 
In addition to substituting the generator architecture, the architecture with a quantum discriminator combined with the classical generator described in MolGAN has been experimented with.
Our goal is to see if the quantum discriminator shows any advantage over its classical counterparts. 
We have found that the number of learnable parameters is significantly less than the classical ones while the model can still generate valid molecules. 
Furthermore, to conduct a fair comparison between the classical discriminator and the quantum one, we changed the MolGAN classical graph-based discriminator to the multiple-layer perceptron (MLP) architecture, which is more similar to our proposed quantum discriminator architecture, reduced the number of its training parameters, and found that quantum discriminator with 50 training parameters outperforms the classical discriminator with 22,500 parameters in terms of generated molecule properties and KL divergence. 
For simplicity, all the combinations of the classical/quantum noise/generator/discriminator are listed in \Cref{tbl:model_name}.
In this experiment, the noises of the generator are always sampled from the Gaussian distribution. 

In the following subsections, we first present our proposed MolGAN-CQ and the training details.
Second, we compare the results of MolGAN-CQ with MolGAN. 
In the end, we made architecture modifications from the Graph-Classical-Discriminator in MolGAN to the MLP-Classical-Discriminator (MolGAN-CC) and reduced the amount of the learnable parameters to make a fair comparison with MolGAN-CQ.

\subsubsection{MolGAN-CQ Architecture and Training Details}
\begin{figure}[hbt!]
    \centering
    \includegraphics[width=0.7\textwidth]{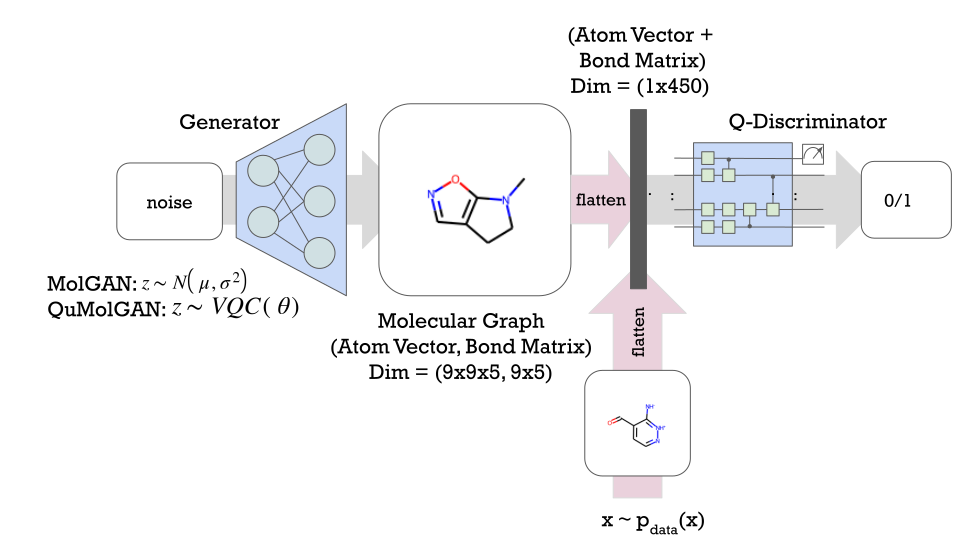}
    %\includesvg[width=0.7\textwidth]{figures/molgan_qdis_schema.svg}
    \caption{Schema of MolGAN-CQ. One of the generator outputs, the bond matrix, is flattened to a vector and concatenated with the other generator outputs, the atom vector, resulting in a vector with the size of 1x450, which is fed into the quantum discriminator.  }
    \label{fig:MolGAN-CQis-schema}
\end{figure}

As illustrated in \Cref{fig:MolGAN-CQis-schema}, the model consists of two components: classical generator and quantum discriminator. 
To encode two generator outputs, a bond matrix, and an atom vector, into the quantum discriminator efficiently, we first flatten the bond matrix into a vector, which is subsequently concatenated with the atom vector to a new vector with the size of 450. 
After that, the resulting vector will be the input of the Q-Discriminator, composed of an amplitude embedding layer followed by 3 strongly entangling layers as described in \Cref{fig:vqc_quantum_discriminator}.

For the training details, at first, we followed the best hyper-parameters set in MolGAN. 
However, applying the same training details on MolGAN-CQ makes the training process hard to converge. 
Therefore, we change the learning rate from $1 \times 10^{-3}$ to $1\times10^{-4}$ for the generator to make it more stable.
Moreover, we have experimented with the alternating training times between the classical generator and quantum discriminator with several sets, including (G, D)=(1, 5), (1, 8), (1, 10). 
We found out that training 1 step C-Generator followed by 10 steps Q-Discriminator stabilizes the training process. 
The other hyperparameters remain identical to the ones in MolGAN.

\subsubsection{Comparison with classical MolGAN}
\label{sec:com_with_molgan}
\begin{figure}[hbt!]
    \centering
    \includegraphics[width=0.7\textwidth]{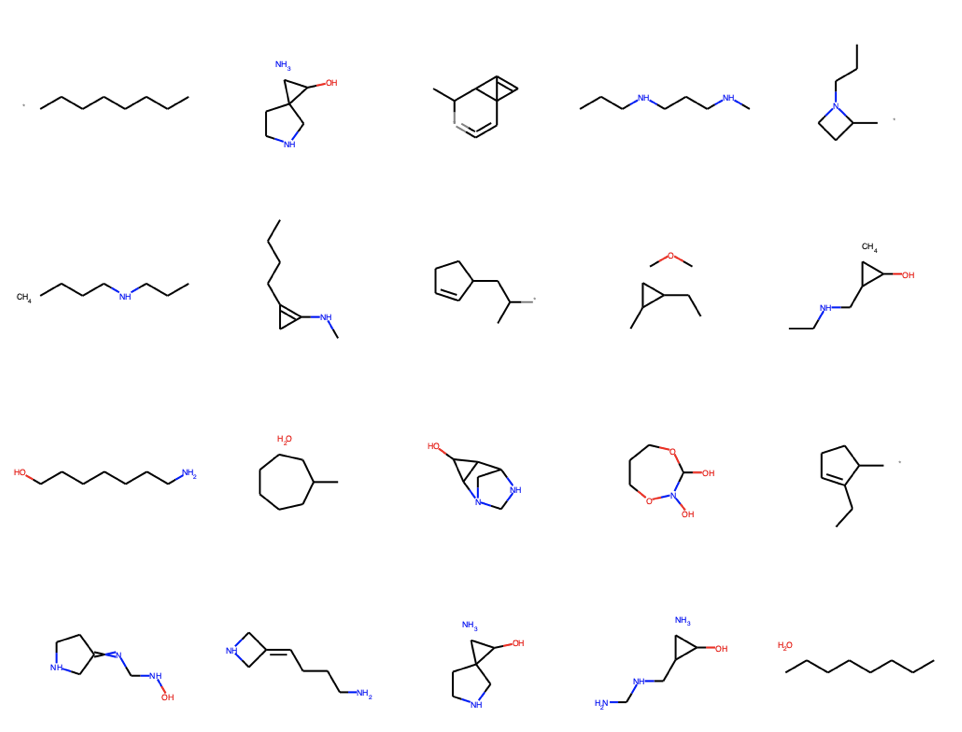}
    \caption{Example molecules from MolGAN-CQ. }
    \label{fig:MolGAN_CQ_example_molecules}
\end{figure}

In this section, we compare MolGAN-CQ with MolGAN. 
We trained MolGAN according to the best hyper-parameter sets found in the previous section. 
Notice here, since training a quantum network is time-consuming, and the loss curve from classical MolGAN at epoch 30 shows the trend of convergence, both MolGAN-CQ and MolGAN were only trained to epoch 30 and evaluated at epoch 30 instead of epoch 150. 
After that, both models generate 5000 samples to do a further comparison. 
\Cref{tbl:molgan_molgan_cq} demonstrates that MolGAN-CQ can generate valid and drug-like molecules.
In addition, MolGAN-CQ can generate molecules with better drug properties, particularly in Solute and SA. 
However, compared to classical MolGAN, MolGAN-CQ does not have an advantage in KL divergence with training data probability.
\Cref{fig:MolGAN_CQ_example_molecules} shows some molecules generated from MolGAN-CQ.

\begin{table}[hbt!]
\caption{Performance comparison between MolGAN and MolGAN-CQ. The models are only trained for 30 epochs. Bold numbers indicate better scores. The QED, Solute, and SA scores in this table are calculated from the valid molecules.}
  \label{tbl:molgan_molgan_cq}
\begin{tabular}{ccc}
\hline
                 & \textbf{MolGAN} & \textbf{MolGAN-CQ} \\ \hline
\# of molecules\textsuperscript{\emph{a}}  & \textbf{2,693}  & 730                 \\ 
QED $\uparrow$             & 0.47            & \textbf{0.48}                \\ 
Solute $\uparrow$          & 0.31            & \textbf{0.44}                \\ 
SA $\uparrow$              & 0.31            & \textbf{0.66}                \\ 
KL Score (S)\textsuperscript{\emph{b}}$\uparrow$     & \textbf{0.94}   & 0.75                \\ \hline
\end{tabular}\\
\textsuperscript{\emph{a}} Number of valid and unique molecules from 5,000 samples;\\
\textsuperscript{\emph{b}} From \Cref{eq:kl}
\end{table}

\subsubsection{Comparison with MolGAN-CC }
\begin{figure}[hbt!] % "[t!]" placement specifier just for this example
\begin{subfigure}{0.5\textwidth}
\includegraphics[width=\linewidth]{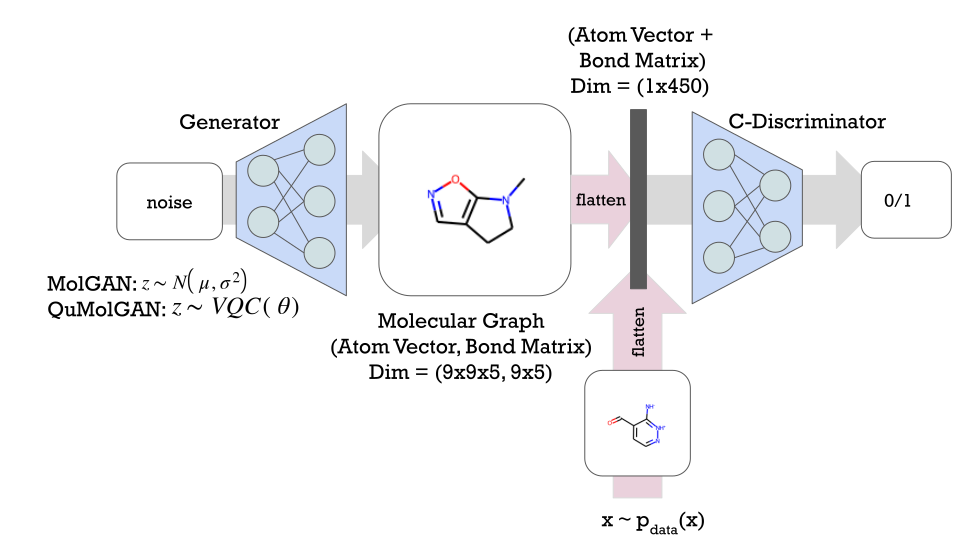}
%\includesvg[width=\linewidth]{figures/molgan_cc_schema.svg}
\caption{Schema of MolGAN-CC} 
\end{subfigure}\hspace*{\fill}
\begin{subfigure}{0.5\textwidth}
\includegraphics[width=\linewidth]{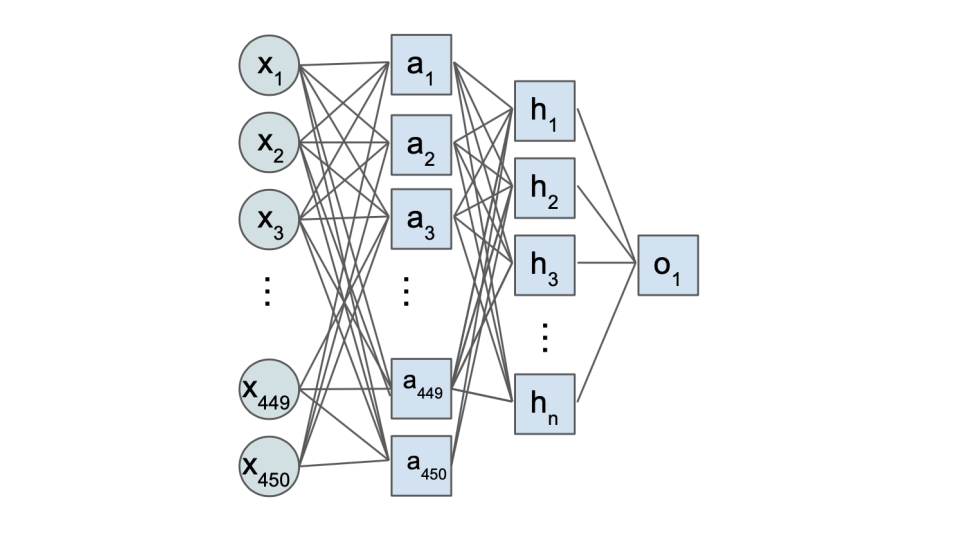}%\includesvg[width=\linewidth]{figures/molgan_cc_dis.svg}
\caption{MLP-based discriminator architecture in MolGAN-CC.}
\end{subfigure}
\caption{(a) Schema of MolGAN-CC. (b) MLP-based discriminator architecture in MolGAN-CC.  The number of the hidden layers and neurons varies in the classical MLP-based discriminator as \Cref{tbl:molgan_cc} }
\label{fig:molgan_cc_schema}
\end{figure}

In this section, MolGAN-CQ is compared with MolGAN-CC with different numbers of hidden layers to evaluate MolGAN-CQ's capacity. 
To have a fair comparison, we modified the original graph-based network to a multiple-layer perceptron (MLP).
Under this condition, the input of the MLP-based discriminator would be the same as the one in MolGAN-CQ's quantum discriminator, a flattened vector instead of a graph. 
The network architecture of MolGAN-CC is shown in \Cref{fig:molgan_cc_schema}. 

Furthermore, since the discriminator of MolGAN-CQ only has 50 learnable parameters, we have tried to reduce the discriminator size of MolGAN-CC as small as possible for a fair and reasonable evaluation. 
However, the input vector is already a size of 450, it is not possible to decrease the parameter size to 50 in the classical MLP. 
Therefore, we have tried MolGAN-CC with 3 different parameter sizes of discriminator as shown in \Cref{tbl:molgan_cc}.

\begin{table}[hbt!]
\caption{The details of MolGAN-CC models with the varied size of discriminators.}
  \label{tbl:molgan_cc}
\begin{tabular}{ccc}
\hline
              & \textbf{\# of layer in the MolGAN-CC discriminator}                                              & \textbf{parameter size} \\ \hline
MolGAN-CC-ER\textsuperscript{\emph{a}} & \begin{tabular}[c]{@{}c@{}}3\\ (450, 50, 1)\end{tabular} & 22K                     \\ 
MolGAN-CC-HR\textsuperscript{\emph{b}} & \begin{tabular}[c]{@{}c@{}}3\\ (450, 100, 1)\end{tabular}         & 45K                     \\ 
MolGAN-CC-NR\textsuperscript{\emph{c}} & \begin{tabular}[c]{@{}c@{}}4 \\ (450, 150, 50, 1)\end{tabular}    & 82K                     \\ \hline
\end{tabular}\\
\textsuperscript{\emph{a}} Extremely-reduction;\\
\textsuperscript{\emph{b}} Highly-reduction
\textsuperscript{\emph{c}} No-reduction;
\end{table}

\begin{figure}[hbt!]
    \centering
    \includegraphics[width=0.8\textwidth]{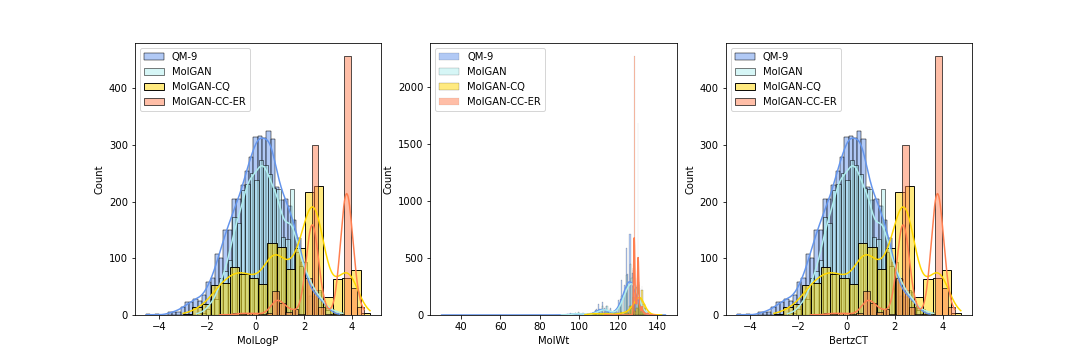}
    \caption{Molecular properties of generated molecules. Generated molecules from MolGAN-CC-ER, MolGAN-CQ, and MolGAN were analyzed in each column with the molecular properties of  MolLogP, MolWt, and BertzCT. In each graph, the histogram with blue indicates the distribution of the training data QM9. }
    \label{fig:molgan_cq_molgan_cc}
\end{figure}

\Cref{tbl:molgan_cq_molgan_cc} demonstrates that although MolGAN-CC-NR and MolGAN-CC-HR have a higher capacity to generate molecules whose molecule properties are more similar to the training data, MolGAN-CQ with only 50 parameters can achieve an outstanding performance compared to MolGAN-CC-ER with 22K parameters in terms of KL-score which shows the quantum advantage in the expression power. The distributions of molecular properties of MolLogP, MolWt, and BertzCT, generated from MolGAN-CC-ER, MolGAN-CQ, MolGAN, and QM9 are shown in \Cref{fig:molgan_cq_molgan_cc}. As we can see from \Cref{fig:molgan_cq_molgan_cc}, molecular properties from MolGAN are closer to the ones from the training data, QM9, compared to MolGAN-CQ and MolGAN-CC-ER. Nevertheless, MolGAN-CQ with only 50 parameters could generate molecules with similar distribution to the training data in comparison to MolGAN-CC-ER with 22K parameters. 

\begin{table}[hbt!]
\caption{Performance comparison between MolGAN-CQ and MolGAN-CC with 3 different sizes of the MLP-based discriminator. Bold numbers indicate better performance across different types of models. The QED, Solute, and SA scores in this table are calculated from the valid molecules.}
\label{tbl:molgan_cq_molgan_cc}
\begin{tabular}{ccccc}
\hline
                 & \textbf{MolGAN-CQ} & \textbf{MolGAN-CC-ER} & \textbf{MolGAN-CC-HR} & \textbf{MolGAN-CC-NR} \\ \hline
\# of parameters & 50                  & 22K                    & 45K                    & 82K                    \\
\#of molecules\textsuperscript{\emph{a}} & \textbf{2,693}      & 104                    & 1919                   & 2284                   \\
QED $\uparrow$              & 0.47                & \textbf{0.51}                   & 0.49                   & 0.5                    \\
Solute $\uparrow$           & 0.44                & \textbf{0.63}                   & 0.35                   & 0.38                   \\
SA $\uparrow$               & 0.66                & \textbf{0.97}                   & 0.48                   & 0.50                   \\
KL Score (S)\textsuperscript{\emph{b}}$\uparrow$     & 0.75                & 0.28                   & \textbf{0.84}          & 0.81          \\ \hline    
\end{tabular}\\
\textsuperscript{\emph{a}} Number of valid and unique molecules from 5,000 samples;\\
\textsuperscript{\emph{b}} From \Cref{eq:kl} \\
\end{table}

\section{Conclusions}

In this paper, we have explored the quantum advantage in small molecule drug discovery by substituting each part of MolGAN \cite{de2018molgan} with a VQC step by step and comparing its performance with the classical counterpart.
In the first experiment, using a VQC as a noise generator to the classical GAN can generate small molecules with better drug properties including QED, SA, and LogP particularly when the input dimension to the generator is small, e.g., z\_dim = 2 or z\_dim = 3.
However, QuMolGAN has difficulty generating as many unique molecules compared to classical MolGAN which results in a lower KL Score. 
In addition, this hybrid model achieves better performance in the goal-directed benchmark compared to the classical counterpart. 
In the second experiment, we substitute the classical generator with a VQC with the patch method \cite{huang2021experimental}. We demonstrated the potential of generating training-set-like small molecules using a quantum generator.
However, the training processing is resource-consuming and time-consuming even in the advanced classical computer.
In the third experiment, we replace the classical discriminator with a VQC and compare its performance with the classical counterpart. 
This hybrid model MolGAN-CQ outperforms the classical counterpart in terms of generated molecule properties and the KL score. 
We also demonstrated that the hybrid model could generate valid molecules with only tens of learnable parameters in a quantum discriminator.
\mh{The proposed hybrid model has the potential to be integrated into the Insilico Medicine Chemistry42\textsuperscript{TM} \cite{ivanenkov2021chemistry42} platform (\Cref{fig:chemistry42})}.

\bibliographystyle{apsrev4-2}
\bibliography{refs.bib}

\newpage
\appendix
\section{Different Generator Complexities} \label{appendix:generator_complexity}

We would like to explore the relationship between the generator complexity and the training data. 
The generator of MolGAN consists of multiple dense layers (linear layer followed by activation layer and dropout layer \cite{srivastava2014dropout}). 
We train three models with different generator complexities. 
The generator of the first model has the same amount of parameters as the original MolGAN \cite{de2018molgan}.
The generator of the second model only has approximately 15\% of the parameters as the original MolGAN, and we call this model MolGAN-HR (high reduction).
The generator of the third model merely has approximately 2\% of the parameters as the original MolGAN, and we call this model MolGAN-ER (extreme reduction).   
These three models are trained using the same hyperparameters. 
The batch size is 128. 
The models are trained for 150 epochs. 
The learning rate is 0.001 for both the generator and discriminator. 
The discriminator is updated three times followed by updating the generator once. 
We evaluate each model using the generator from 150 epochs, and the results are shown in \Cref{tbl:g_conv_dim}.
In this table, MolGAN-HR can generate the most valid and \textbf{\textit{unique}} molecules which results in the highest KL Score.  
The QED, Solute, and SA scores in this table are calculated from the valid molecules.

\begin{table}[hbt!]
  \caption{Different complexities of generator in MolGAN. QED: a quantitative estimate of drug-likeness. SA: synthesizability. The QED, Solute, and SA scores in this table are calculated from the valid molecules.}
  \label{tbl:g_conv_dim}
  \resizebox{\textwidth}{!}{%
  \begin{tabular}{cccccccccc}
    \toprule 
    Model &  \# molecules\textsuperscript{\emph{d}} & validity $\uparrow$ & uniqueness $\uparrow$ & novelty $\uparrow$ & diversity $\uparrow$ & QED $\uparrow$ & Solubility $\uparrow$ & SA $\uparrow$ & KL Score (S)\textsuperscript{\emph{e}} $\uparrow$ \\
    \midrule
    MolGAN\textsuperscript{\emph{a}} & 995 & 78.54 & 25.34 & 59.56 & 0.57  & 0.47 & 0.31 & 0.28 & 0.893 \\
    MolGAN-HR\textsuperscript{\emph{b}} & 3191 & 78.34 & 81.47 & 64.33 & 0.59 & 0.47 & 0.31 & 0.29 & 0.954 \\
    MolGAN-ER\textsuperscript{\emph{c}} & 2826 & 75.84 & 74.53 & 67.14 & 0.67 & 0.47 & 0.31 & 0.32 & 0.945 \\
    \bottomrule
  \end{tabular}
  }
  \textsuperscript{\emph{a}} 396,610 parameters; \\
  \textsuperscript{\emph{b}} Highly-reduction (59,202 parameters);\\
  \textsuperscript{\emph{c}} Extremely-reduction (7,794 parameters); \\
  \textsuperscript{\emph{d}} Number of valid and unique molecules from 5,000 samples; \\ 
  \textsuperscript{\emph{e}} From \Cref{eq:kl}; \\
\end{table}

\section{Varying the Input Noise Dimension of Generator} \label{appendix:noise_dimension}

Padala et al \cite{padala2021effect} have examined the effect of input noise dimension in GANs in generating the Gaussian distribution, the handwritten digit recognition \cite{lecun1998gradient}, and the celebrity image classification \cite{liu2015deep}. 
They observed a significant effect on the results when the input noise dimension is alternated.
They also claimed that the optimal noise dimension is depended on the dataset and loss function used. 
Inspired by their work, we would like to examine the optimal input noise dimension for the task of small molecular generation using the QM9 dataset for MolGAN and QuMolGAN.
The input noise dimension is varied from 1 to 8 for the MolGAN-HR, and the results are shown in \Cref{tbl:z_dim_molgan}.
For QuMolGAN-HR, we only examine the ansatz with 2, 3, 4, and 8 qubits, and the experimental results are shown in \Cref{tbl:q_quantum_molgan_mr}.
These models are trained for 150 epochs. 
The learning rates of the discriminator and generator are 0.001, and the learning rate of the quantum circuit is 0.04. 
The QED, Solute, and SA scores in this table are calculated from the valid molecules.
From these two tables, the number of valid and \textit{\textbf{unique}} molecules saturates when the input dimension (z\_dim) goes beyond 4. 
The KL Score of MolGAN-HR oscillates between 0.94 and 0.96, and the KL Score of QuMolGAN-HR has the best KL Score as z\_dim=4. 

\begin{table}
  \caption{Different input noise dimensions of generator in MolGAN-HR. The QED, Solute, and SA scores in this table are calculated from the valid molecules.}
  \label{tbl:z_dim_molgan}
  \resizebox{\textwidth}{!}{%
  \begin{tabular}{cccccccccc}
    \toprule 
    z-dim & \# molecules\textsuperscript{\emph{a}} & validity $\uparrow$ & uniqueness $\uparrow$ & novelty $\uparrow$ & diversity $\uparrow$ & QED $\uparrow$ & Solubility $\uparrow$ & SA $\uparrow$ & KL Score (S)\textsuperscript{\emph{b}} $\uparrow$\\
    \midrule
    8 & 3204 & 80.04 & 80.06 & 64.67 & 0.62 & 0.47 & 0.31 & 0.30 & 0.946 \\
    7 & 3206 & 80.42 & 79.73 & 65.61 & 0.74 & 0.47 & 0.30 & 0.30 & 0.957 \\
    6 & 2898 & 77.76 & 74.54 & 61.34 & 0.72 & 0.47 & 0.30 & 0.29 & 0.937 \\
    5 & 3104 & 78.86 & 78.72 & 63.78 & 0.62 & 0.47 & 0.30 & 0.30 & 0.955 \\
    4 & 2930 & 81.26 & 72.11 & 63.82 & 0.65 & 0.47 & 0.30 & 0.30 & 0.941 \\
    3 & 1850 & 75.36 & 49.10 & 66.43 & 0.68 & 0.47 & 0.31 & 0.30 & 0.926 \\
    2 & 571  & 73.56 & 15.52 & 72.92 & 0.66 & 0.48 & 0.32 & 0.34 & 0.868 \\
    1 & 33   & 59.16 & 1.12  & 77.45 & 0.72 & 0.47 & 0.33 & 0.29 & 0.687\\
    \bottomrule
  \end{tabular}
  }
  \textsuperscript{\emph{a}} Number of valid and unique molecules from 5,000 samples; \\
  \textsuperscript{\emph{b}} From \Cref{eq:kl}; \\
  
\end{table}

\begin{table}
  \caption{Different numbers of qubits in the quantum circuit of QuMolGAN-HR. The QED, Solute, and SA scores in this table are calculated from the valid molecules.}
  \label{tbl:q_quantum_molgan_mr}
  \resizebox{\textwidth}{!}{%
  \begin{tabular}{cccccccccc}
    \toprule 
    \# qubits\textsuperscript{\emph{a}}  & \# molecules\textsuperscript{\emph{b}} & validity $\uparrow$ & uniqueness $\uparrow$ & novelty $\uparrow$ & diversity $\uparrow$ & QED $\uparrow$ & Solubility $\uparrow$ & SA $\uparrow$ & KL Score (S)\textsuperscript{\emph{c}} $\uparrow$ \\
    \midrule
    8 & 515 & 73.94 & 13.93 & 60.05 & 0.62 & 0.46 & 0.31 & 0.27 & 0.785\\
    4 & 534 & 70.02 & 15.25 & 71.04 & 0.70 & 0.47 & 0.32 & 0.29 & 0.894\\
    3 & 301 & 64.74 & 9.30 & 66.23 & 0.58 & 0.50 & 0.34 & 0.38 & 0.782 \\
    2 & 194 & 63.68 & 6.09 & 66.02 & 0.69 & 0.49 & 0.36 & 0.45 & 0.570 \\
    \bottomrule
  \end{tabular}
  }
  \textsuperscript{\emph{a}} Number of qubits of quantum circuit; \\
  \textsuperscript{\emph{b}} Number of valid and unique molecules from 5,000 samples; \\
  \textsuperscript{\emph{c}} From \Cref{eq:kl}; \\
\end{table}

\section{Number of Parameterized Layers of Variational Quantum Circuit} \label{appendix:parameterized_layers}

Variational quantum circuits (VQCs) can be considered machine learning models with remarkable expressive power for a variety of data-driven tasks, such as supervised learning and generative modeling \cite{benedetti2019parameterized}.
VQCs consists of three ingredients: a fixed initial state, e.g., the zero state, a quantum circuit, parameterized by a set of learnable parameters, and the measurement as shown in \Cref{fig:quantum_circuit_noise_generator}.
A quantum circuit is composed of an initialization layer and parameterized layer(s).
The initialization layer brings randomness to the VQC, and the parameters of parameterized layer(s) can be learned through back-propagation. 
One challenge in implementing variational quantum algorithms is to choose an effective circuit that well represents the solution space while maintaining a low circuit depth and the number of parameters \cite{sim2019expressibility}.
Increasing the number of parameterized layers of VQC has the potential to improve the impressive power of VQC and its performance.
However, it may face the problem of Barren plateaus \cite{mcclean2018barren,zhang2021toward,zhang2022escaping} and longer computational time. 
In this experiment, we would like to find the optimal number of parameterized layers that can generate the most unique and valid molecules for QuMolGAN-HR. 
We have examined one to five parameterized layers, and the results are shown in \Cref{tbl:pl_quantum_molgan_mr}.
In this table, QuMolGAN-HR achieves the best KL Score when the VQC has three parameterized layers. 

\begin{table}
  \caption{Different parameterized layers of QuMolGAN-HR. The QED, Solute, and SA scores in this table are calculated from the valid molecules.}
  \label{tbl:pl_quantum_molgan_mr}
  \resizebox{\textwidth}{!}{%
  \begin{tabular}{ccccccccccc}
    \toprule 
    \# layers\textsuperscript{\emph{a}}  & \# molecules\textsuperscript{\emph{b}} & validity $\uparrow$ & uniqueness $\uparrow$ & novelty $\uparrow$ & diversity $\uparrow$ & QED $\uparrow$ & Solubility $\uparrow$ & SA $\uparrow$ & KL Score (S)\textsuperscript{\emph{c}} $\uparrow$ \\
    \midrule
    5 & 375 & 70.10 & 10.70 & 70.44 & 0.63 & 0.47 & 0.32 & 0.31 & 0.738 \\
    4 & 377 & 52.04 & 14.49 & 64.64 & 0.57 & 0.46 & 0.32 & 0.27 & 0.701 \\
    3 & 443 & 72.40 & 12.24 & 69.20 & 0.68 & 0.50 & 0.36 & 0.32 & 0.784 \\
    2 & 435 & 76.06 & 11.44 & 71.00 & 0.61 & 0.45 & 0.33 & 0.28 & 0.765 \\
    1 & 165 & 75.98 & 4.34 & 67.81 & 0.66 & 0.44 & 0.32 & 0.26 & 0.702 \\
    
    \bottomrule
  \end{tabular}
  }
  \textsuperscript{\emph{a}} Number of parameterized layers; \\
  \textsuperscript{\emph{b}} Number of valid and unique molecules from 5,000 samples; \\
  \textsuperscript{\emph{c}} From \Cref{eq:kl}; \\
\end{table}

\section{Chemistry42\textsuperscript{TM}}

\begin{figure}[hbt!]
    \centering
    \includegraphics[width=0.39\textwidth]{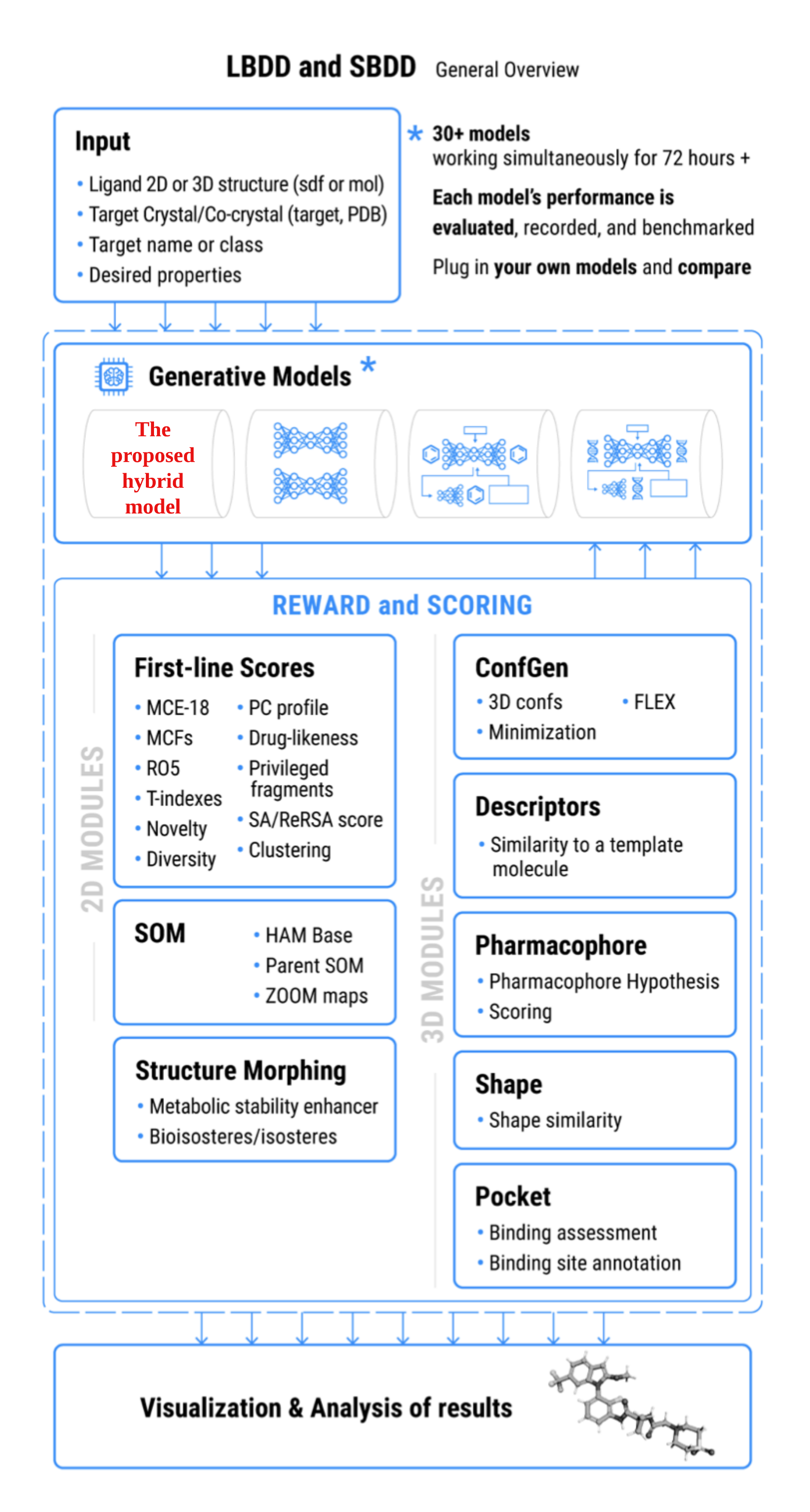}
    \caption{
    Overall Workflow of Insilico Medicine Chemistry42\textsuperscript{TM} platform. The proposed hybrid generative model can be integrated into the Generative Models section of the platform. More details of Chemistry42\textsuperscript{TM} can be found in \cite{ivanenkov2021chemistry42}.
    }
    \label{fig:chemistry42}
\end{figure}

\end{document}